\documentclass{emulateapj}
\usepackage{url}
\usepackage{gensymb}
\usepackage{textcomp}
\usepackage{xspace}
\usepackage{mathtools}
\usepackage{natbib,amsmath} \usepackage{graphicx}
\usepackage{hyperref} \usepackage{float} \usepackage{subfigure}
\usepackage{tabularx}
\providecommand{\e}[1]{\ensuremath{\times 10^{#1}}}
\newcommand{\um}{\textmu m\xspace}

\newcommand{\platon}{\texttt{PLATON}\xspace}
\usepackage{array}
\newcolumntype{C}{>{$}c<{$}} 
\newcolumntype{R}{>{$}r<{$}} 

\DeclarePairedDelimiter\abs{\lvert}{\rvert}%

\begin{document}
\title{\platon II: New Capabilities and a Comprehensive Retrieval on HD 189733b Transit and Eclipse Data}

\author{ Michael Zhang\altaffilmark{1}, Yayaati Chachan\altaffilmark{2},
  Eliza M.-R. Kempton\altaffilmark{3,4}, Heather A. Knutson\altaffilmark{2}, Wenjun (Happy) Chang\altaffilmark{5} }

\altaffiltext{1}{Department of Astronomy, 
	California Institute of Technology, Pasadena, CA 91125, USA; 
	mzzhang2014@gmail.com}
\altaffiltext{2}{Division of Geological and Planetary Sciences,
  California Institute of Technology, Pasadena, CA 91125, USA}
\altaffiltext{3}{Department of Astronomy, University of Maryland, College Park, MD 20742, USA}
\altaffiltext{4}{Department of Physics, Grinnell College, 1116 8th Avenue, Grinnell, IA 50112, USA}
\altaffiltext{5}{California Institute of Technology, Pasadena, CA 91125, USA}

\begin{abstract}
Recently, we introduced PLanetary Atmospheric Tool for Observer Noobs (\platon), a Python package that calculates model transmission spectra for exoplanets and retrieves atmospheric characteristics based on observed spectra.  We now expand its capabilities to include the ability to compute secondary eclipse depths.  We have also added the option to calculate models using the correlated-$k$ method for radiative transfer, which improves accuracy without sacrificing speed.  Additionally, we update the opacities in \platon--many of which were generated using old or proprietary line lists--using the most recent and complete public line lists.  These opacities are made available at R=1000 and R=10,000 over the $0.3-30$ \um range, and at R=375,000 in select near IR bands, making it possible to utilize \platon for ground-based high resolution cross correlation studies.  To demonstrate \platon's new capabilities, we perform a retrieval on published \emph{HST} and \emph{Spitzer} transmission and emission spectra of the archetypal hot Jupiter HD 189733b.  This is the first joint transit and secondary eclipse retrieval for this planet in the literature, as well as the most comprehensive set of both transit and eclipse data assembled for a retrieval to date.  We find that these high signal-to-noise data are well-matched by atmosphere models with a C/O ratio of $0.66_{-0.09}^{+0.05}$ and a metallicity of $12_{-5}^{+8}$ times solar where the terminator is dominated by extended nanometer-sized haze particles at optical wavelengths.  These are among the smallest uncertainties reported to date for an exoplanet, demonstrating both the power and the limitations of HST and Spitzer exoplanet observations.
\end{abstract}

\section{Introduction}
The emission spectra of exoplanets provide unique insights into their atmospheric properties \citep[e.g.,][]{madhusudhan_2018}.  By measuring the difference between the in-eclipse and out-of-eclipse flux when the planet passes behind the star (`secondary eclipse'), one can measure the flux emitted by the planet as a function of wavelength.  Secondary eclipse observations probe the compositions and temperature-pressure profiles of their dayside atmospheres.  This technique was used to derive the first atmospheric composition measurement for a Neptune-mass planet (GJ 436b; \citealt{stevenson_2010}), the first definitive detection of a thermal inversion in the atmosphere of an ultra-hot Jupiter \citep{haynes_2015}, and dayside water abundance measurements for several hot Jupiters (e.g. \citealt{kreidberg_2014,line_2016,pinhas_2019}).  Emission spectroscopy at high spectral resolution ($R > 20,000$) has led to the detection of CO, H$_2$O, and HCN in exoplanet atmospheres (i.e. \citealt{snellen_2010,birkby_2018}), including in non-transiting planets like 51 Pegasi b \citep{brogi_2013}. By measuring the atmospheric compositions of these planets, we can obtain new insights into present-day atmospheric processes such as disequilibrium chemistry (e.g., \citealt{moses_2013}), as well as their past formation and migration histories (e.g., \citealt{oberg_2011,madhusudhan_2014,ali-dib_2017,cridland_2019,booth_2019}).

Ideally, we would extract atmospheric parameters from observed emission spectra using a Bayesian retrieval code.  However, there is an overall lack of open source retrieval codes that can handle emission spectra.  A similar lack of retrieval codes for transit spectra motivated us to write \platon \citep{zhang_2019}, a fast, open source, easy to use, and easy to understand forward modeling and retrieval code that traces its lineage back to Exo-Transmit \citep{kempton_2017}.  \platon has since been used in several papers: a few exploring the atmospheric properties of observed planets \citep{chachan_2019,kirk_2019,guo_2020}, and one demonstrating the possibility of using K-means clustering to speed up retrievals by 40\% \citep{hayes_2020}.  In the latter study, the speed of \platon was especially useful due to the necessity of running many retrievals.  We now expand \platon's capabilities to include thermal emission and compare the resulting models and atmospheric retrievals to that of another retrieval code in order to validate this new functionality.

We utilize the open source TauREx code \citep{waldmann_2015b,al-refaie_2019}, which has been used in multiple published studies (i.e. \citealt{komacek_2019, shulyak_2019}), for this comparison.  Specially, we compare to TauREx 3, the latest release.  \platon and TauREx were developed independently of each other, and there are several key differences between their functionalities.  TauREx is a sophisticated code which supports free retrieval of chemical abundances in addition to equilibrium chemistry retrievals.  \platon only allows for retrievals using equilibrium chemistry, with atmospheric metallicity and C/O ratio as free parameters.  When calculating this equilibrium chemistry, \platon uses GGchem, which can account for losses due to condensation \citep{woitke_2018}, whereas TauREx assumes that everything stays in the gas phase.  \platon also uses opacities generated from the latest line lists for water \citep{polyansky_2018}, methane \citep{rey_2017}, and ammonia \citep{coles_2019}, which are significantly more complete and accurate than the line lists that were available at the time TauREx was released.  

The two codes also differ in their treatment of aerosols.  Both support Rayleigh scattering, although \platon also supports Rayleigh-like scattering by allowing for a variable scattering strength and slope ($\sigma(\lambda) = A\sigma_{\rm Rayleigh}(\lambda) (1 \mu m/\lambda)^s$).  Both also support Mie scattering, but with different parameterizations.  TauREx has three ways of approximating Mie opacity: a gray opacity, the parameterization of \cite{lee_2013} ($Q_{ext} = \frac{5}{Q_0 x^{-4} + x^0.2}$), and an analytical calculation of the Mie opacity of spherical particles with a size distribution given by Equation 36 or Equation 37 of \cite{sharp_2007}.  \platon analytically calculates the Mie opacity of spherical particles with a lognormal size distribution, as explained in \cite{zhang_2019}.   In this study we explore the impact of these differences on our models using the benchmark hot Jupiter HD 189733b \citep{bouchy_2005} as our test case. 

Aside from TauREx, other open source retrieval codes include the recently released Helios-r2 \citep{kitzmann_2020} and the Bayesian Atmospheric Radiative Transfer (BART) code \citep{blecic_2016b, blecic_2017}.  Helios-r2 is primarily intended for brown dwarfs and supports both free retrievals and gas-only equilibrium retrievals without clouds.  It does not use a parameterized T/P profile, but retrieves the temperatures of individual layers in the atmosphere, with constraints on how much the temperature can vary from layer to layer.  This approach is suitable for the high signal-to-noise regime of brown dwarfs, but not ideal for exoplanets.

BART is partially described in one subsection of a dissertation \citep{blecic_2016b} and one subsection of a paper \citep{blecic_2017}, but has not been described in detail in a peer reviewed paper.  BART uses a custom MCMC code, MC$^3$, to perform retrievals using either free abundances or gas-only equilibrium chemistry, the latter of which is computed by TEA \citep{blecic_2016a}.  It supports two T/P profile parameterizations: \cite{madhusudhan_2009} and \cite{guillot_2010}. However, BART does not provide opacity data, and the opacity calculator it provides does not support ExoMol line lists.  In addition, TEA is slow, taking 2-3 s of CPU time per temperature/pressure point, and has convergence problems below 400 K \citep{woitke_2018}.

Last in our roundup of retrieval codes is petitRADTRANS, a forward modelling code that does not support retrievals, but is fast enough to be wrapped in a Bayesian retrieval framework \citep{molliere_2019}.  \cite{molliere_2019} benchmark against \platon and find that although \platon is much faster overall, the two codes are comparable in speed for the same number of wavelength points.  petitRADTRANS only supports free abundances, and uses either correlated-k (R=1000) or line-by-line (R=$10^6$) radiative transfer, with a T/P profile parameterization that is a variant of \cite{guillot_2010}.  To our knowledge, \platon is unique among open source retrieval codes in supporting equilibrium condensation in a Bayesian retrieval framework, and our opacities are based on the most up-to-date line lists.

We utilize \platon to carry out the first joint retrieval on published emission and transmission spectroscopy for HD 189733b, resulting in improved constraints on its atmospheric composition.  \platon is also capable of calculating transit and eclipse spectra at R=375,000, and we compare our model to published high-resolution ($R \sim 100,000$) CRIRES emission spectroscopy for this planet to search for previously reported signatures of H$_2$O and HCN \citep{de_kok_2013,birkby_2013,brogi_2016,cabot_2019} in addition to an unreported molecule: CH$_4$.

In Section \ref{sec:eclipse_calc}, we describe the emission spectrum and secondary eclipse depth calculator.  Section \ref{sec:opacity} describes the opacity update, while Section \ref{sec:improvements} describes other new features and improvements in \platon, including the new correlated-$k$ capability and optional model parameters, including a chemical quench pressure and a \emph{HST}/WFC3 offset.  In Section \ref{sec:hd189733b}, we perform a joint retrieval of the transit and eclipse spectra of HD 189733b to infer its atmospheric properties and confirm published high-resolution detections of H$_2$O while calling into question the reported high-resolution detection of HCN.  We summarize our conclusions in Section \ref{sec:conclusion}.

\section{Emission Spectrum and Secondary Eclipse depth calculator}
\label{sec:eclipse_calc}
\subsection{Algorithm}
Our emission spectrum and secondary eclipse depth calculator utilizes much of the same code as the transit depth calculator presented in our first paper \citep{zhang_2019}.  Given a planetary mass, radius, metallicity, C/O ratio, and temperature-pressure profile, we compute equilibrium molecular abundances for 250 pressures uniformly distributed in log(P).  We then solve the hydrostatic equation to determine the height corresponding to each pressure.  With molecular abundances, temperatures, and pressures as a function of height, the emergent flux is given by:

\begin{align}
    F_{p\lambda} = 2\pi \int_0^\infty \int_0^1 B_{\lambda}(\tau_\lambda) e^{-\tau_\lambda/\mu} d\mu d\tau_\lambda
    \label{eq:emergent_flux}
\end{align}
where $B_{\lambda}(\tau_\lambda)$ is the Planck function at an optical depth of $\tau_\lambda$, and $\mu$ is the cosine of the viewing angle with respect to the vertical.  Here, we are making the assumption that the source function is the Planck function, which in turn requires that scattering contributes negligibly to the emission.  Adding scattering as an emission source would make the problem much more complex because it would be non-local: the source function at a certain location would depend on scattered photons from other locations.

Directly integrating Equation \ref{eq:emergent_flux} would require evaluating the integrand hundreds of millions of times--once for every combination of wavelength, $\tau_\lambda$, and $\mu$.  We instead rewrite the double integral as a single integral with a special function:

\begin{align}
     F_{p\lambda} = 2\pi \int_0^\infty B_{\lambda}(\tau_\lambda) E_2(\tau_\lambda) d\tau_\lambda
     \label{eq:emergent_flux_E2}
\end{align}
where $E_2$ is the exponential integral, defined as:

\begin{align}
    E_2(x) = \int_1^\infty \frac{e^{-xt}}{t^2} dt
\end{align}

We discretize and reformulate the equation as follows:
\begin{align}
    F_{p\lambda} &\approx 2\pi\sum_0^N B(\tau_{\lambda, i}) E_2(\tau_{\lambda,i}) \Delta\tau_{\lambda,i}\\
    &\approx 2\pi\sum_0^{N-1} -B(\tau_{\lambda, i}) (E_3(\tau_{\lambda,i+1}) - E_3(\tau_{\lambda,i})),\\
    \label{eq:E3}
\end{align}

where the second equation follows from the first because the integral of $E_2(x)$ is $-E_3(x) + C$.  Reformulating the equation in this way has the virtue of guaranteeing that the result is exactly correct for an isothermal atmosphere.  For an isothermal atmosphere, B is constant; therefore, for every layer i except the two boundaries, the $E_3(\tau_{i+1})$ term is cancelled by the $-E_3(\tau_i)$ term in layer i+1.  The numerical integration therefore gives $2\pi B (E_3(\tau_{\lambda, 0}) - E_3(\tau_{\lambda, N}))$.  Assuming the top of the atmosphere has an optical depth of 0 and the bottom has an optical depth of infinity, we obtain $F = \pi B$--exactly the correct result for an isothermal atmosphere.  In practice, we find that this trick reduces the error for non-isothermal atmospheres as well, especially when the temperature gradient is small.  This method of replacing $g(x) \Delta x$ with $G(x_2) - G(x_1)$ (where g(x) is the derivative of G(x)) before integrating was inspired by the TauREx source code, which uses the same technique in combination with Gaussian quadrature.

\begin{figure}[ht]
  \centering 
  \includegraphics[width=\linewidth]{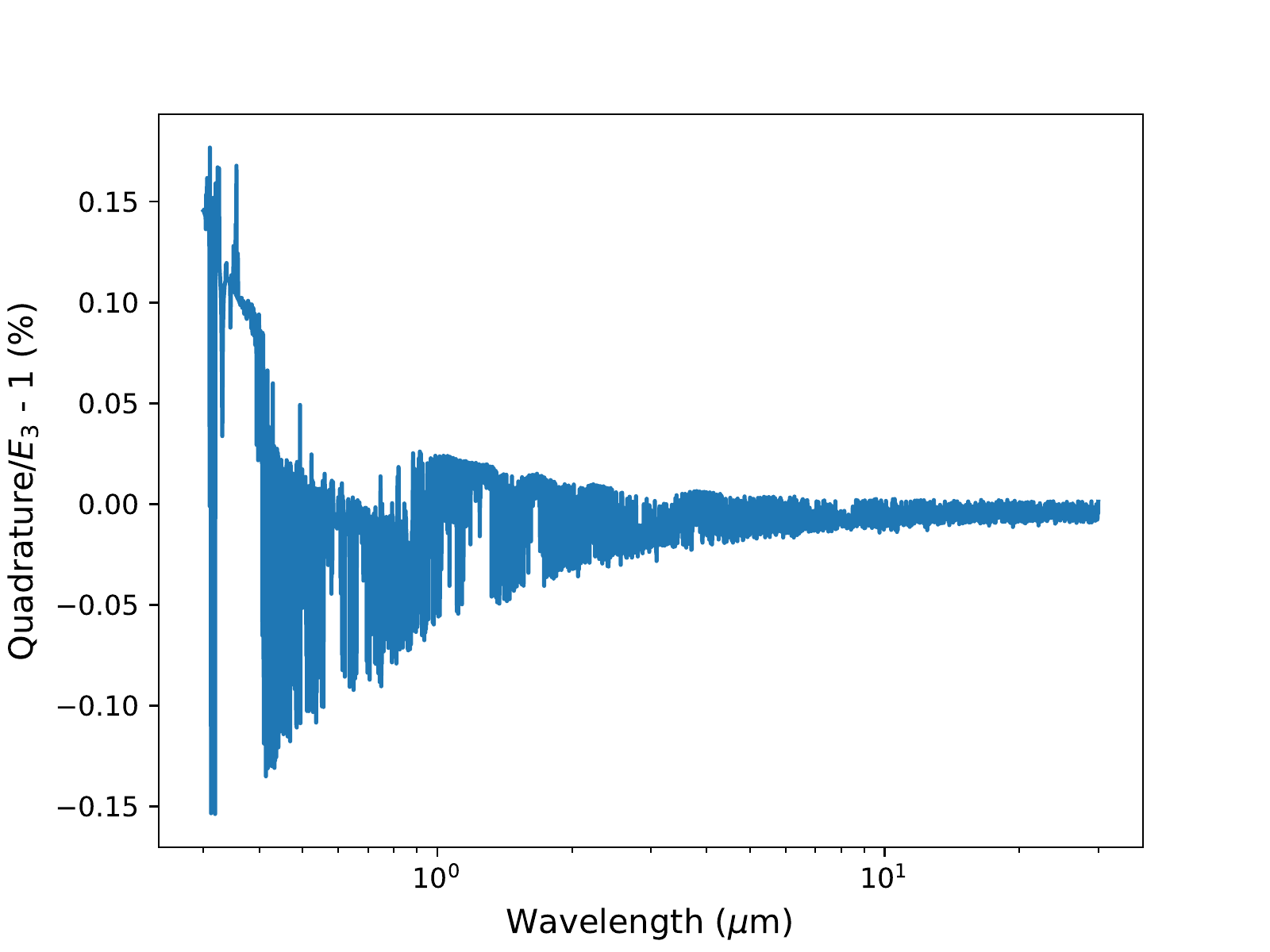}
  \caption{Fractional error introduced by using Gaussian quadrature with 4 points instead of $E_3$, for the test planet in Subsection \ref{subsec:validation}.}
\label{fig:quadrature_vs_E3}
\end{figure}

The $E_3$ exponential integral is a special function defined by scipy, eliminating the need to perform integrals to evaluate it.  It is also continuous, infinitely differentiable, and approaches 0.5 as $x \to 0$ and 0 as $x \to \infty$.  These properties mean that $E_3$ poses no problems for numerical integration.  To further speed up the code, we evaluate $E_3(x)$ on a logarithmic grid spanning $x=10^{-6}$ to $x=10^2$ during initialization of the eclipse depth calculator, and interpolate from this grid thereafter.  The interpolation is accurate to $1.2 \times 10^{-5}$, and is therefore a negligible source of error.  The more common approach to integrating Equation \ref{eq:emergent_flux}, adopted by TauREx and HELIOS-r2, is to use Gaussian quadrature.  Compared to our approach, Gaussian quadrature is 4 times slower and introduces errors of $\sim$0.2\% for the test planet in Subsection \ref{subsec:validation} when 4 points are used (Figure \ref{fig:quadrature_vs_E3}).  This error is utterly negligible compared to the other sources of error we explore in Subsection \ref{subsec:validation}.

To derive a monochromatic eclipse depth from the emergent flux, we multiply by the square of the planet-to-star radius ratio and divide by the stellar emergent flux:

\begin{align}
    D_\lambda = \bigg(\frac{R_{p\lambda}}{R_s}\bigg)^2 \frac{F_{p\lambda}}{F_{s\lambda}}
    \label{eq:eclipse_depth}
\end{align}
The stellar spectrum is calculated by interpolating the BT-Settl (AGSS2009) stellar spectral grid \citep{allard_2012}, as provided by the Spanish
Virtual Observatory\footnote{\url{http://svo2.cab.inta-csic.es/theory/newov2/index.php}}.  Here, the wavelength-dependent planet radius $R_{p\lambda}$ is defined as the radius at which the radial optical depth reaches one, assuming that the limb has the same temperature-pressure profile as the dayside.  (This is not to be confused with the white light planet radius the users pass into \platon, which is the radius at a pressure of 1 bar.)  This approach differs from those of previous studies (i.e. \citealt{waldmann_2015b}), which typically fix the radius to the value measured from the optical or near-infrared transit depth.  This is larger than the effective planet size for emission spectroscopy, as starlight transmitted through the planet's limb should reach an optical depth of one at lower pressures than for direct emission.  Our approach, although an improvement, is still only an approximation.  The planetary limb is usually colder than the day side, and even if it were not, Equation \ref{eq:eclipse_depth} is only strictly true if the atmospheric scale height is a negligible fraction of the planetary radius.  These inaccuracies are expected to cause errors on the order of 1\% for hot Jupiters, since the scale height of a hot Jupiter is of order 1\% of the radius.  For further discussion of the difficulty in choosing a photospheric radius and the error this introduces, we refer the reader to \cite{fortney_2019}.

When fitting observational data, we typically want band-integrated fluxes and secondary eclipse depths.  The band-integrated flux is equal to the number of photons emitted from the planet within the band, and the secondary eclipse depth is that value divided by the number of photons emitted from the star within the band.  The eclipse depth for the band $\lambda_1$--$\lambda_2$ is then:

\begin{align}
    D = \frac{\int_{\lambda_1}^{\lambda_2} R_{p\lambda}^2 F_{p\lambda} \lambda d\lambda}{\int_{\lambda_1}^{\lambda_2} R_s^2 F_{s\lambda} \lambda d\lambda}
    \label{eq:band_depth}
\end{align}
We utilize Equation \ref{eq:band_depth} when the user defines custom wavelength bins.  Otherwise, we compute monochromatic fluxes and secondary eclipse depths at full spectral resolution (R=1000) using Equation \ref{eq:eclipse_depth}.

\subsection{Temperature-pressure profile}
The temperature-pressure profile is a crucial component of any atmospheric emission model.  This profile determines whether molecular features will be seen in absorption or emission and their relative strengths.  Although it is possible to predict the T/P profile theoretically using energy balance arguments, such a self-consistent calculation is computationally expensive.  This is because the radiative intensity, opacity, chemical abundance, and temperature at each pressure level all depend on one another, requiring an iterative procedure to solve for all components simultaneously.  These iterative procedures are too slow for our purposes here.  We instead forgo self-consistency and retrieve a parameterized T/P profile along with the atmospheric composition. We support three parametric forms for the T/P profile, listed in order of increasing complexity: isothermal, \cite{line_2013}, and \cite{madhusudhan_2009}.

\subsubsection{Isothermal}
An isothermal profile always gives a blackbody planetary spectrum.  This can be derived theoretically from Equation \ref{eq:emergent_flux}.

\subsubsection{Line et al. (2013)}
This physically motivated parameterization was invented by \cite{guillot_2010} to shed light on exoplanet atmospheres, and then subsequently extended by  \cite{line_2013}.  \cite{guillot_2010} used the two-stream approximation with radiation partitioned into two distinct wavelength channels: thermal and visible.  Starlight is considered purely visible, while planetary emission is considered purely thermal.  The planetary atmosphere is assumed to have a single opacity $\kappa_{th}$ that applies to thermal radiation everywhere, and a single opacity $\kappa_v$ that applies to visible radiation everywhere.  Under these simplifying assumptions, the averaged dayside T/P profile can be derived analytically:

\begin{align}
    \tau &= \frac{\kappa_{th}P}{g}\\
    m(\gamma) &\equiv 1 + \frac{1}{\gamma} [1 + (\frac{\gamma \tau}{2} - 1)e^{-\gamma \tau}] + \gamma(1-\frac{\tau^2}{2})E_2(\gamma \tau)]\\
    T^4 &= \frac{3}{4}T_{int}^4 (\frac{2}{3} + \tau) + \frac{1}{2}T_{eq}^4 m(\gamma)
    \label{eq:line_TP_profile}
\end{align}
where T is the temperature at a certain height in the atmosphere, $\tau$ is the optical depth of thermal radiation from the top of the atmosphere corresponding to that height, $\gamma \equiv \kappa_v/\kappa_{th}$, $T_{eq} \equiv T_* \sqrt{\frac{R_*}{2a}}$ is the equilibrium temperature assuming zero albedo, $E_2$ is the exponential integral with n=2, and $T_{int}$ is a temperature reflecting the amount of internal heat.

\cite{line_2013} introduce an albedo into this formulation, so that the equilibrium temperature is now $T_{eq} \equiv \beta T_* \sqrt{\frac{R_*}{2a}}$.  They also introduce a second visible channel with its own opacity to allow for temperature inversions, so that the temperature is now:

\begin{align}
    T^4 &= \frac{3}{4}T_{int}^4 (\frac{2}{3} + \tau) + (1-\alpha)\eta(\gamma_1) + \alpha \eta(\gamma_2)
\end{align}
where $\gamma_1 \equiv \kappa_{v1}/\kappa_{th}$,  $\gamma_2 \equiv \kappa_{v2}/\kappa_{th}$, and $\alpha$ partitions the visible radiation between the two channels.  Unlike \cite{line_2013}, we impose the constraint that $\alpha <= 0.5$: that is, the second visible stream is by definition the minor one.  Without this constraint, the two streams are interchangeable.

In total, this parameterization has six free parameters: $\kappa_{th}$, $\gamma_1$, $\gamma_2$, $\alpha$, $\beta$, and $T_{int}$.  Following \cite{line_2013}, we recommend fixing $T_{int}$ to 100 K in most cases because internal heat usually contributes negligibly to the short-period transiting planets that are most amenable to atmospheric characterization using the secondary eclipse technique.  Exceptions to this rule may include eccentric planets, planets whose cooling is delayed, or planets whose radii are inflated. 

\subsubsection{Madhusudhan \& Seager (2009)}
\cite{madhusudhan_2009} introduced a purely empirical T/P profile for exoplanet atmosphere modeling, which was designed to be flexible enough to approximate most published theoretical T/P profiles from forward models without having an excessive number of free parameters.  This model divides the atmosphere into three layers: a deep isothermal layer (caused by the limited interior flux compared to the stellar flux, as $dT/dr \propto L(r)$ in a radiative zone where the diffusion approximation holds), an intermediate layer that can support a thermal inversion, and an outer layer, intended to represent the optically thin region. The parametric profile is agnostic about physical assumptions (e.g. convection, optical depths) that set the temperature and pressure structure of the atmosphere.  The temperatures of the three layers are then:

\begin{align*}
    T_{outer} = T_0 + \frac{ln(P/P_0)^2}{\alpha_1^2}\\
    T_{mid} = T_2 + \frac{ln(P/P_2)^2}{\alpha_2^2}\\
    T_{inner} = T_3\\
\end{align*}
There are six free parameters in this model: $T_0$, $P_1$, $\alpha_1$, $\alpha_2$, $P_3$, $T_3$.  Following \cite{madhusudhan_2009}, we set $P_0$ to the pressure at the top of the atmosphere, which for us is $10^{-4}$ Pa.  $T_2$ and $P_2$ are set by the requirement that temperature must be continuous across region boundaries:

\begin{align*}
    ln(P_2) &= \frac{\alpha_2^2 (T_0 + ln(P_1/P_0)^2/\alpha^2 - T_3) - ln(P_1)^2 + ln(P_3)^2}{2ln(P_3/P_1)}\\
    T_2 &= T_3 - \frac{ln(P_3/P_2)^2}{\alpha_2^2}
\end{align*}

\subsection{Benchmarking, speed advice}
One of the goals of \platon is to be fast.  To illustrate typical speeds, We benchmark \platon on a typical desktop computer to illustrate its performance.  The computer runs Ubuntu 16.04 LTS with a Core i7 7700k CPU and 16 GB of RAM.

\begin{table}[ht]
  \centering
  \caption{Benchmarks for desktop computation of emission forward model}
  \begin{tabular}{c c c c}
  \hline
  	Band & $\lambda$ ($\mu m$) & Time (ktables/R=1k/R=10k)\\
   \hline
    All wavelengths & 0.3--30 & 0.26/0.24/4.8 s\\
    WFC3 & 1.119--1.628 & 0.029/0.029/0.26 s\\
    Spitzer 3.6 $\mu m$ & 3.2--4.0 & 0.021/0.021/0.11 s\\
    Spitzer 4.5 $\mu m$ & 4.0--5.0 & 0.020/0.021/0.11 s\\
      \hline
  \end{tabular}
  \label{table:forward_model_benchmarks}
\end{table}

When the forward model is first initialized, \platon loads all relevant data files into memory.  This takes 0.34 s (1.6 s for R=10,000), but is only done once.  Table \ref{table:forward_model_benchmarks} shows the amount of time taken to compute eclipse depths within the most commonly used bands once \platon is initialized.  The time taken  depends linearly on the number of wavelength grid points within the band.  Since grid points are spaced uniformly in logarithmic space, the number of grid points is proportional to the ratio between the maximum and minimum wavelengths.  The time taken also depends approximately linearly on the resolution, for the same reason.  Because our correlated k algorithm runs at R=100 with 10 Gaussian quadrature points, it performs the same number of radiative transfer computations as the R=1000 opacity sampling method, explaining the very similar running times.

It is difficult to give a representative running time for nested sampling retrievals, because this is highly dependent on the problem at hand.  The running time is proportional to the total logarithmic wavelength range, the number of live points used, and the log of the ratio between the prior parameter hypervolume and the posterior hypervolume.  The hypervolume ratio depends on the width of the priors, and on the quality of the data: an exquisite dataset takes longer to retrieve on.

Despite these variations, some rough numbers are possible.  With 1000 live points, generously wide priors, and the exquisite HD 189733b dataset, \texttt{dynesty} required 400,000 likelihood evaluations for the eclipse-only retrieval.  A typical retrieval with a lower signal-to-noise data set and a less conservative prior range would require fewer likelihood evaluations; using 200 live points instead of 1000 would cut the number of evaluations by a factor of 5.  Taking 200,000 evaluations as a typical value for 1000 live points, we see that retrieving on a dataset of WFC3, Spitzer 3.6 $\mu m$, and Spitzer 4.5 $\mu m$ observations will take 0.8 hours with 200 live points and R=1000 opacities (or correlated k coefficients); 4 hours with 1000 live points and R=1000 opacities; 5.3 hours with 200 live points and R=10,000 opacities; and 27 hours with 1000 live points and R=10,000 opacities.

We recommend a staged approach to retrievals.  Exploratory data analysis can be done with R=1000 opacities and 200 live points.  In the process, intermittent spot checks should be performed with R=10,000 opacities and 200 live points to check the effect of resolution, and with R=1000 opacities and 1000 live points to check the effect of sparse sampling.  When one is satisfied with the exploratory data analysis and is ready to finalize the results, one should run a final retrieval with R=10,000 opacities and 1000 live points.  This is the approach we followed for HD 189733b, although had we stuck with the low-resolution, sparsely sampled retrieval, none of our conclusions would have changed.

If these running times are still too slow, there is one trivial way to speed up the code by a factor of a few: by going to PLATON\_DIR/data/Absorption and removing the absorption files of all molecules that have a negligible effect on the spectrum, which prevents \platon from taking their opacities into account.  For a hot Jupiter, for example, the vast majority of the molecules in \platon (see Table \ref{table:line_list_sources} for a list) are unimportant.  One might reasonably include CO, CO$_2$, CH$_4$, H$_2$O, NH$_3$, H$_2$S, and HCN in an emission retrieval, but neglect the other 23.  This decreases the R=10k running time for the entire 0.3--30 \um range from 4.8 s to 2.1 s.  In the near future, we will implement an opacity zeroer in \platon that implements this functionality without having to touch the data files.

\subsection{Validation}
\label{subsec:validation}
To validate \platon's new emission spectroscopy mode, we test two cases: an isothermal atmosphere and a non-isothermal atmosphere modelled on that of HD 189733b.  In the case of an isothermal atmosphere, the planetary flux should be equal to that of a blackbody:

\begin{align}
    F_\lambda = \pi B_\lambda(T)
\label{eq:blackbody_flux}    
\end{align}

We find that this is indeed the case.  The numerically evaluated flux differs from the theoretical expectation by an amount consistent with machine precision ($\sim 2^{-52} \approx 2 \times 10^{-16}$).  This is not a surprise, as our numerical integration algorithm (Equation \ref{eq:E3}) gives exactly the correct answer for the special case of an isothermal atmosphere.

\begin{figure}[ht]
  \centering \subfigure {\includegraphics
    [width=0.5\textwidth]{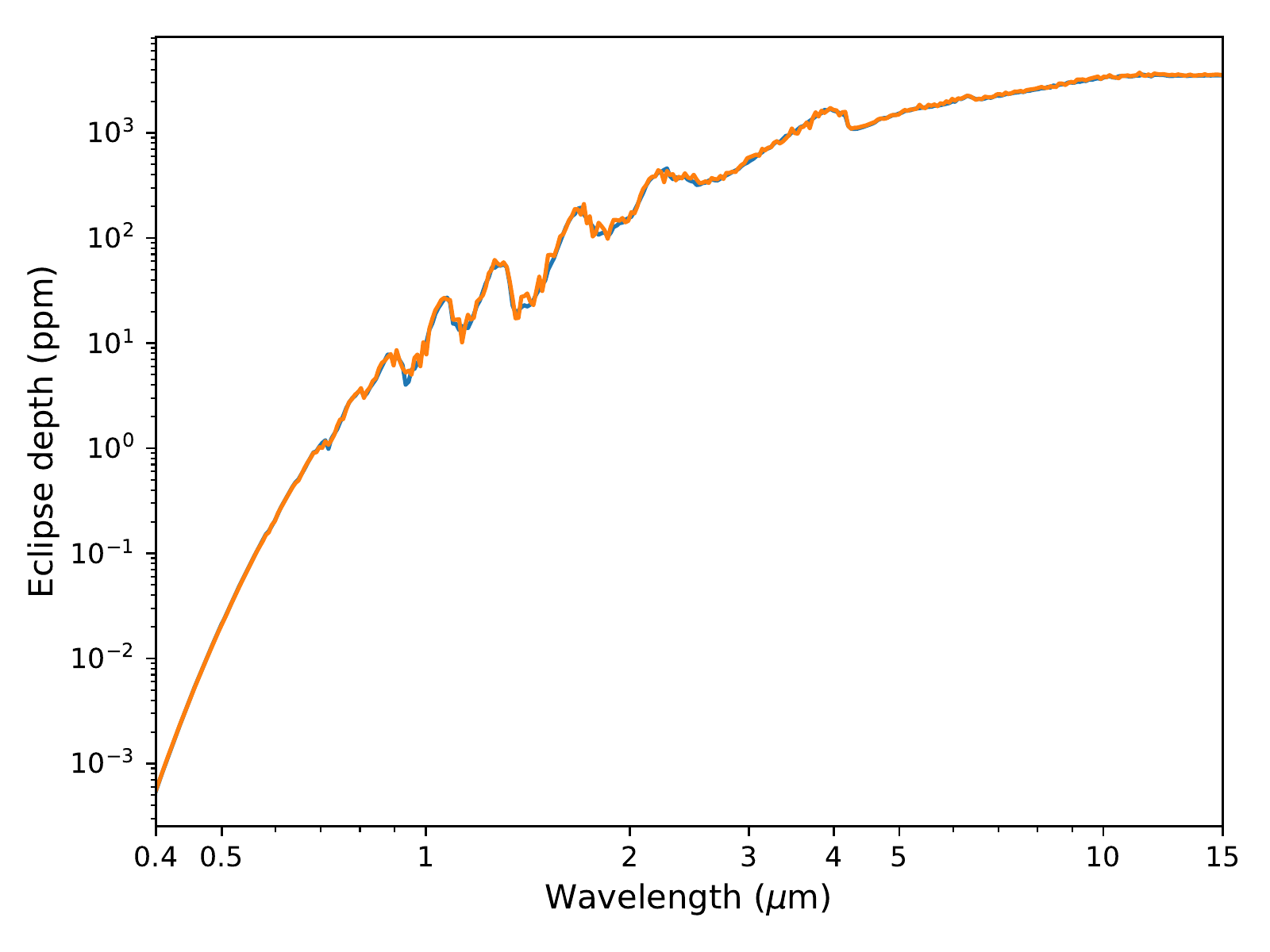}}\qquad
  \subfigure {\includegraphics
    [width=0.5\textwidth]{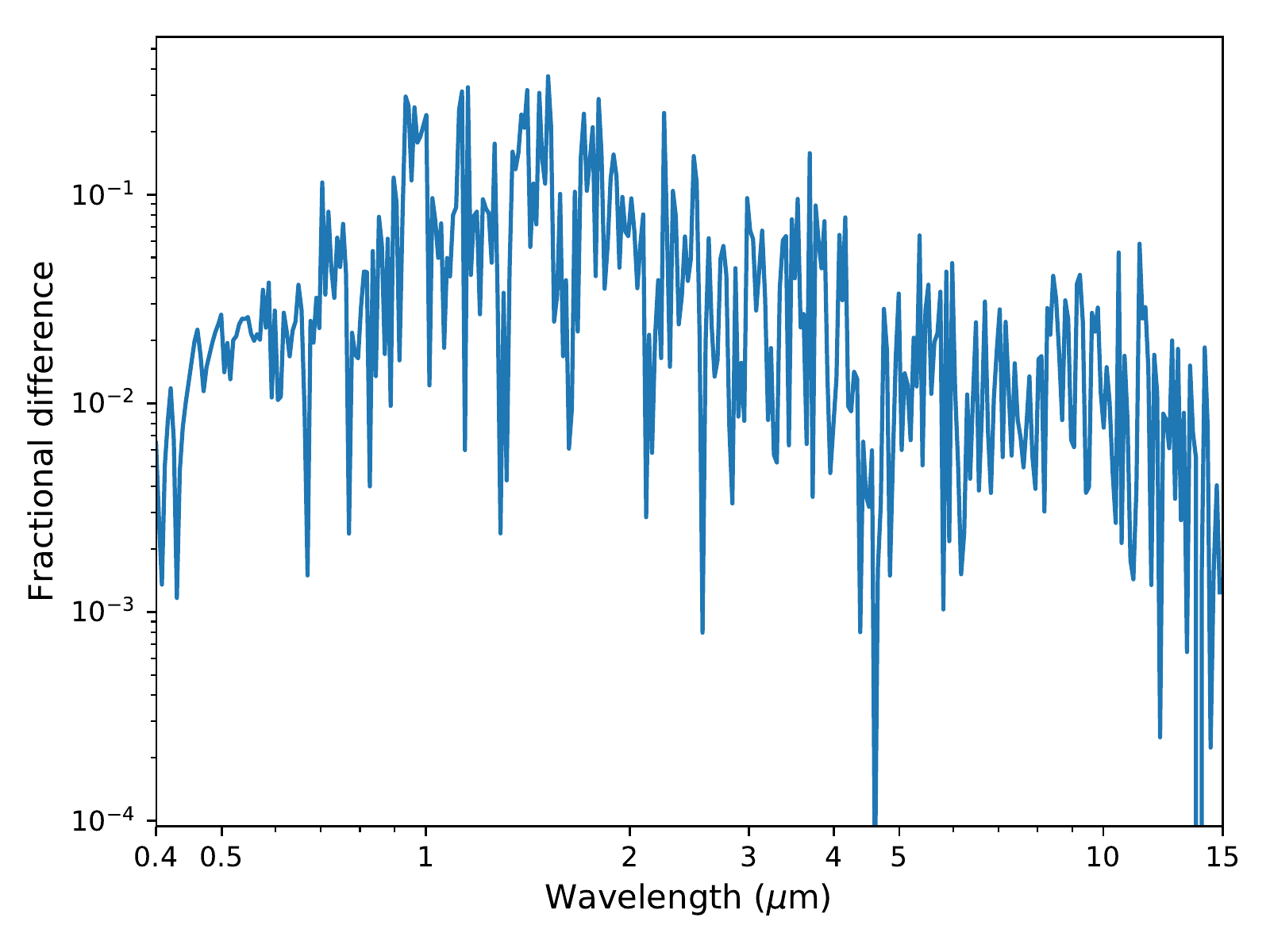}}
    \caption{Top: Comparison of eclipse depths computed by \platon and Tau-REx for a planet meant to approximate HD 189733b.  Bottom: fractional differences between the two models.}
\label{fig:eclipse_depth_comp}
\end{figure}

To validate the non-isothermal atmosphere case, we compared the output of \platon to that of TauREx 3 \citep{al-refaie_2019}.  Using both codes, we simulated a planet meant to represent HD 189733b.  This test planet has the parameters given in Table \ref{table:test_planet_params}.  These parameters are, respectively, the stellar temperature $T_s$, stellar radius $R_s$, planetary mass $M_p$, planetary radius $R_p$, planetary atmospheric metallicity $Z$ relative to the Sun, planetary C/O ratio, planetary equilibrium temperature $T_{eq}$, and the five parameters ($\beta$, $\kappa_{th}$, $\kappa_{v1}$, $\kappa_{v2}$, and $\alpha$) specifying the T/P profile following the formulation of \cite{line_2013}.  Because the chemical equilibrium model of TauREx does not include condensation, we passed include\_condensation=False to \platon, causing \platon to also use a gas-only chemical equilibrium model.  TauREx is distributed with opacities for CO$_2$, NH$_3$, CH$_4$, CO, and H$_2$O; we therefore zero out the abundances of all other active gases in \platon for the purposes of this test.  In addition, we used a blackbody as the stellar spectrum in both codes, as the PHOENIX spectra that TauREx supports do not extend redward of 5 \um.  We generated an emission spectrum from both codes, both binned to R=100 from the native resolution of 1000 for \platon and 15,000 for TauREx, and compared the resulting wavelength-dependent eclipse depths.  As shown in Figure \ref{fig:eclipse_depth_comp}, the median absolute difference between the two is 2.1\%, with a 95th percentile of 17\% and a maximum of 39\%.

\begin{table}[ht]
  \centering
  \caption{Parameters of test planet}
  \begin{tabular}{c c}
  \hline
  	Parameter & Value\\
      \hline
      $T_s$ & 5052 K\\
      $R_s$ & 0.751 $R_{\odot}$\\
      $M_p$ & 1.129 $M_J$ \\
      $R_p$ & 1.144 $R_J$\\
      $Z_p/Z_{\odot}$   &   20\\
      C/O &   0.7\\
      $T_{eq}$ & 1189 K\\
      $\beta$ & 1\\
      $\kappa_{th}$ & 3.8\e{-3} m$^2$ kg$^{-1}$\\
      $\kappa_{v1}$ & 1.9\e{-3} m$^2$ kg$^{-1}$\\
      $\kappa_{v2}$ & 5.2\e{-4} m$^2$ kg$^{-1}$\\
      $\alpha$ & 0.331\\
      \hline
  \end{tabular}
  \tablecomments{This test planet is chosen to have properties broadly similar to those of HD 189733b.  $T_s$, $R_s$, and $M_p$ are taken from \citealt{strassun_2017}, and we selected values for the other parameters that approximately reproduce HD 189733b's transit and eclipse spectra.}
  \label{table:test_planet_params}
\end{table}

\begin{figure}[ht]
  \centering \subfigure {\includegraphics
    [width=0.5\textwidth]{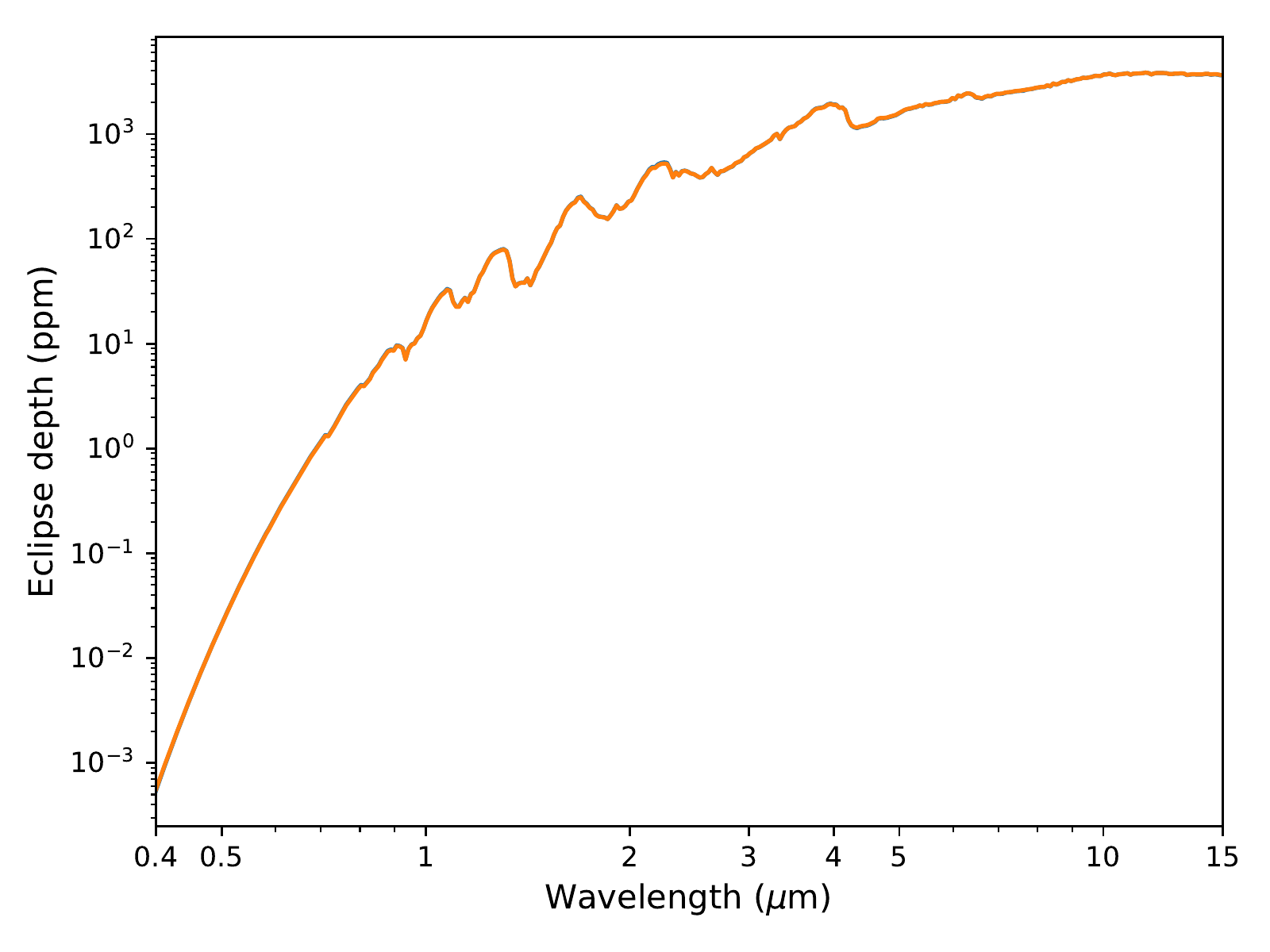}}\qquad
  \subfigure {\includegraphics
    [width=0.5\textwidth]{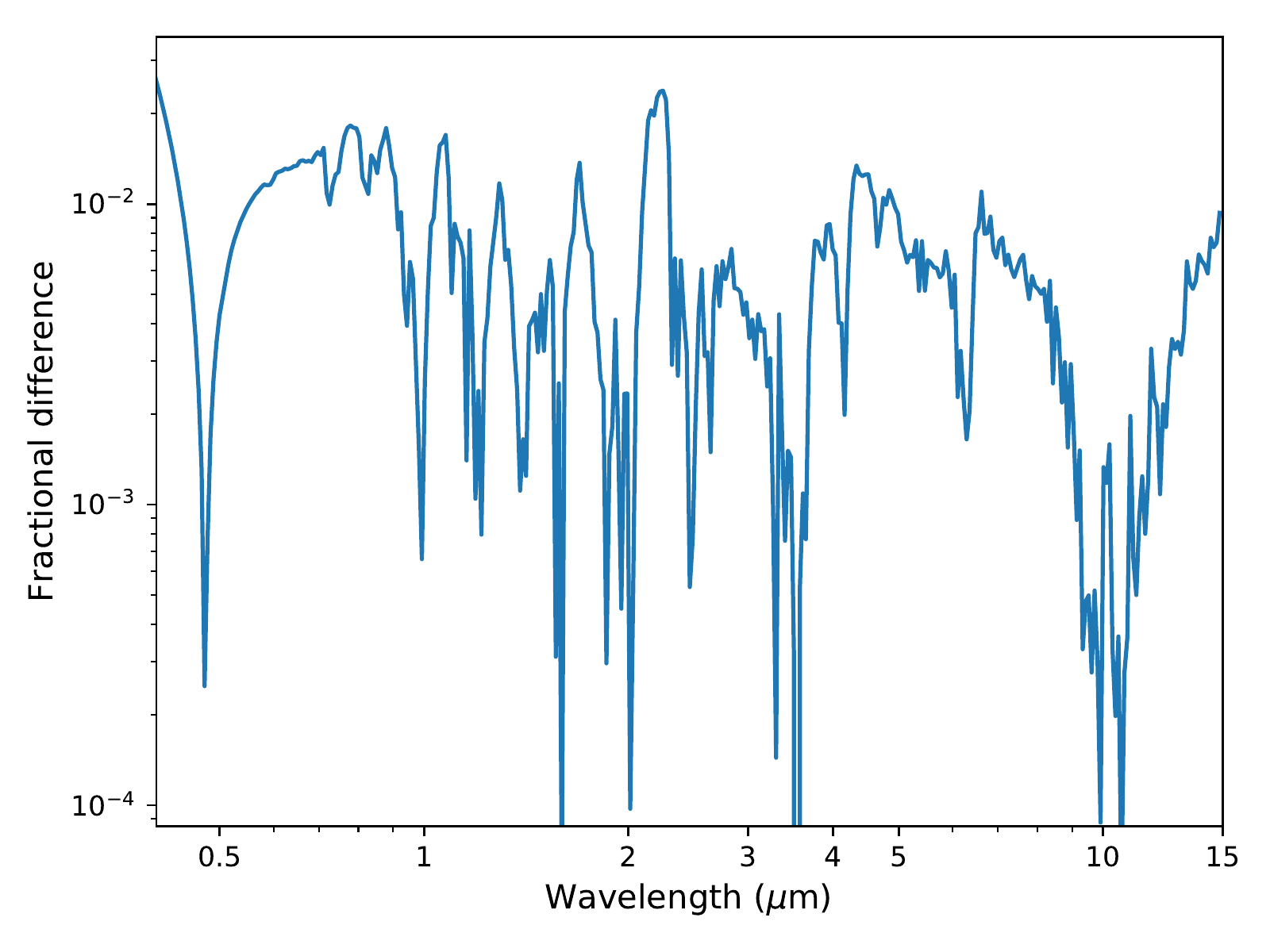}}
    \caption{Comparison of eclipse depths computed by \platon and Tau-REx (top) and corresponding residuals (bottom), where both codes are fed the same mixing ratios and the same opacities at the same resolution of R=10,000.}
\label{fig:eclipse_depth_comp_final}
\end{figure}

\begin{figure}[ht]
  \centering
  \subfigure {\includegraphics
    [width=0.5\textwidth]{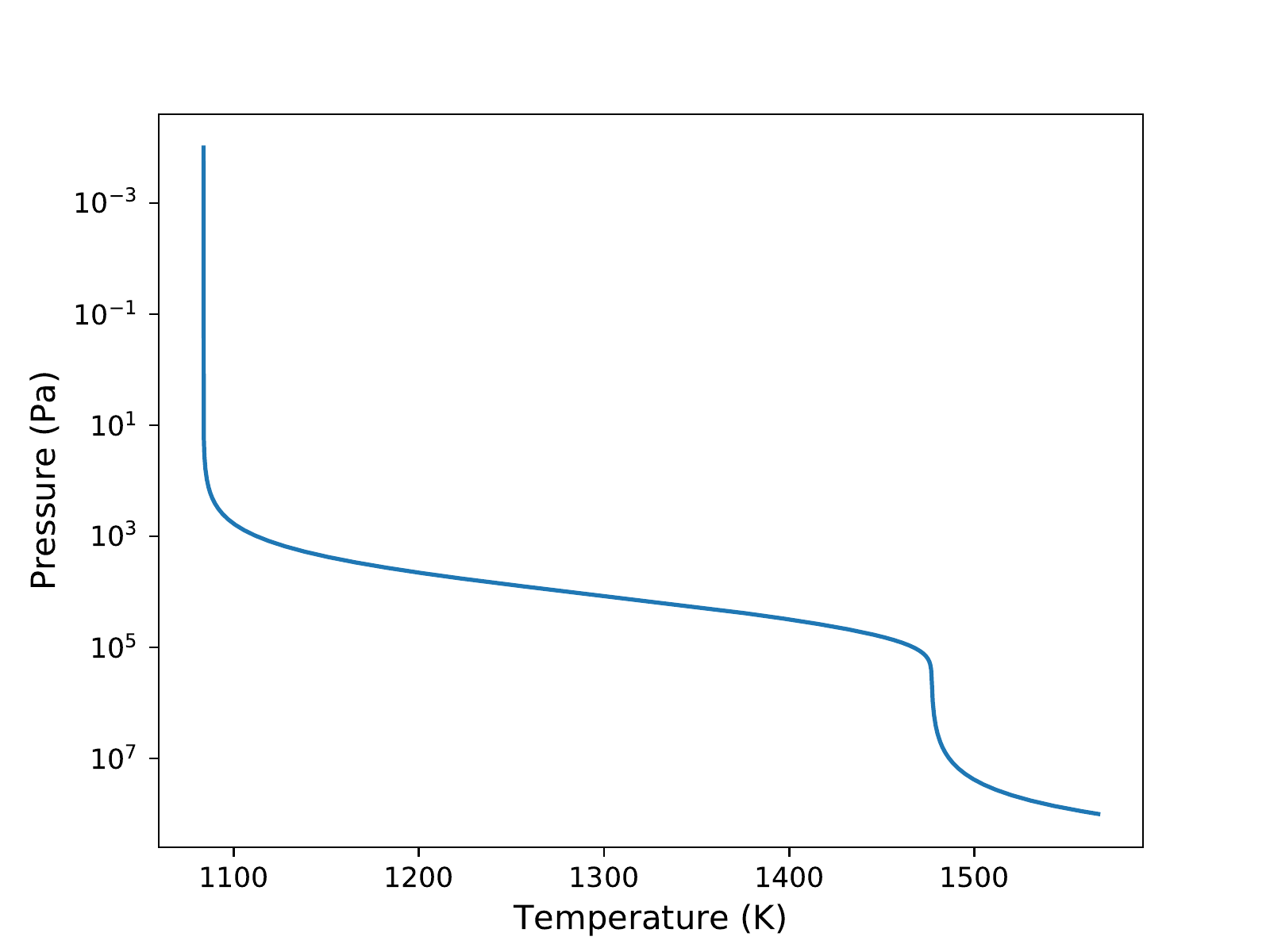}}\qquad\subfigure {\includegraphics[width=\linewidth]{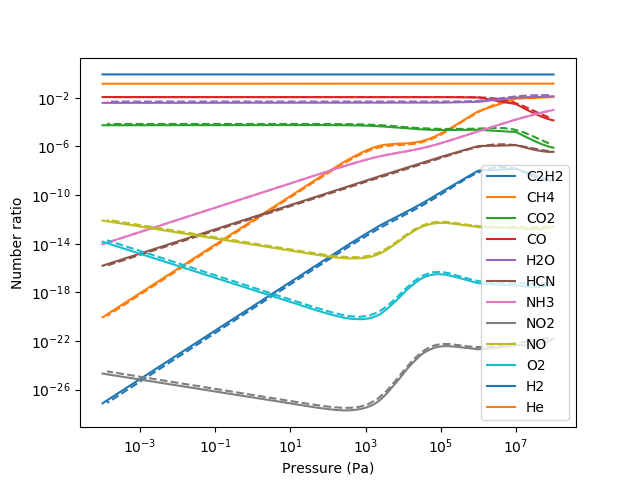}}
  \caption{Differences in molecular abundances under equilibrium chemistry conditions between GGchem (used by \platon) and ACE (used by Tau-REx) for the test planet, with the T/P profile shown in the upper panel.  These differences are typically tens of percent.}
\label{fig:platon_taurex_composition_comp}
\end{figure}

There are a number of differences between \platon and Tau-REx that could explain the discrepancy in the predicted eclipse depths.  First, \platon performs radiative transfer at a spectral resolution of R=1000, while we ran Tau-REx with R=15,000 opacity files.  Second, the two codes also handle equilibrium chemistry differently.  Tau-REx uses ACE \citep{agundez_2012} and only considers the elements H, He, C, O, and N and 106 molecules composed of those elements, while \platon also includes F, Na, Mg, Si, P, S, Cl, K, Ti, and V, along with 300 molecules composed of those elements.  This results in abundance differences in the tens of percent, as seen in Figure \ref{fig:platon_taurex_composition_comp}.  Third, \platon also uses a wavelength-dependent radius to convert emergent flux to luminosity (Equation \ref{eq:eclipse_depth}), but Tau-REx does not.  Fourth, \platon's opacities are generated from newer line lists than the opacities currently available from Tau-REx's website.  Finally, there are slight differences in the algorithm used to calculate the planet-star flux ratio.  For example, in calculating the emergent flux (Equation \ref{eq:emergent_flux}), Tau-REx evaluates the integral over viewing angles by sampling four viewing angles and using Gaussian quadrature, whereas \platon evaluates the integral analytically, which is equivalent to using an infinite number of viewing angles.

\begin{table*}[ht]
  \centering
  \caption{Effect of removing differences between \platon and TauREx}
  \begin{tabular}{|p{5.3cm}|p{2.2cm}|p{3.4cm}|p{2cm}|}
  \hline
  	Removed differences & Median diff. (\%) & 95th percentile diff. (\%) & Max diff. (\%) \\
      \hline
      None & 2.2 & 19 & 37\\
      Resolution & 1.2 & 7.6 & 22\\
      Resolution, Chemistry & 1.2 & 4.6 & 11\\
      Resolution, Chemistry, Opacities & 0.6 & 1.7 & 2.7\\
      Chemistry, Opacities & 1.9 & 17 & 30\\
      Chemistry, Opacities with ktables & 2.1 & 8.1 & 20\\
      \hline
  \end{tabular}
  \tablecomments{This table shows the discrepancies between eclipse depths calculated by \platon and TauREx after binning to R=100, for a wavelength range of 0.4--15 \um. As differences between the two codes are eliminated, their eclipse depths become more and more similar, as expected.}
  \label{table:platon_taurex_diffs}
\end{table*}

To disentangle which factors cause most of the differences, we modified \platon step by step to more closely approximate Tau-REx's algorithm, re-measuring the discrepancies at each step.  The results are summarized in Table \ref{table:platon_taurex_diffs}, and described in detail below.  First, we replaced the default R=1000 opacities with R=10,000 opacities, which are also publicly available.  This reduced the median absolute difference between Tau-REx and \platon from 2.1\% to 1.6\%, with the 95th percentile at 7.4\% and some wavelengths having a discrepancy of up to 17\%.  We conclude that a higher resolution leads to significant improvement in the agreement between \platon and Tau-REx at some, but not most, wavelengths.  Next, we disabled equilibrium chemistry and set constant abundances with altitude for CH$_4$, CO$_2$, CO, H$_2$O, NH$_3$, H$_2$, and He.  This results in a median difference of 1.1\% with a 95th percentile of 5.1\% and maximum discrepancies of up to 10\%.  Following this, we replaced the line absorption cross sections in Tau-REx with those used in \platon.  This decreased the median difference to 0.8\%, the 95th percentile to 2.1\%, and the maximum difference from 10\% to 2.7\%.  The spectra produced by the two codes are compared in Figure \ref{fig:eclipse_depth_comp_final}.  The remaining discrepancies are likely due to slight differences in the radiative transfer code, especially the number of viewing angles (4 in Tau-REx vs. infinite in \platon) and the precise interpolation method for absorption cross sections. As a final step, we replaced the R=10,000 opacities with the default R=1000 opacities and redid the comparison.  This time, we obtained a median difference of 1.9\%, a 95th percentile of 15\%, and a maximum of 31\%.  If we do the test with correlated k radiative transfer instead of opacity sampling, these numbers are 2.2\%, 7.4\%, and 19\%.

Based on these tests, we concluded that differences in resolution, chemistry and opacities are all significant contributors to the discrepancies between the two codes.  As shown in Table \ref{table:platon_taurex_diffs}, the median error caused by these differences are on the order of 2\%, with the 95th percentile being 10-20\%, and the maximum difference being a few tens of percent.

\subsection{Retrieval comparison}
To test the validity of \platon retrievals, we performed an equilibrium chemistry retrieval comparison between \platon and TauREx using a synthetic spectrum.  We used TauREx to generate the 0.4--6 $\mu m$ emission spectrum using the stellar and planetary parameters in Table \ref{table:test_planet_params}.  The spectrum is binned down to a resolution of R=100, and 100 ppm of white noise is added to every binned eclipse depth.  We then ran a retrieval on the synthetic spectrum using both TauREx and \platon, with 6 free parameters (corresponding uniform priors, all generously wide, in brackets): $\log{Z}$ (-1--3), C/O (0.2--2.0), $\log{\kappa_{th}}$ (-3.4 -- -1.4), $\log{\gamma_{v_1}}$ (-1 -- 1), $\log{\gamma_{v_2}}$ (-1 -- 1), and $\alpha$ (0--0.5).  The \platon retrieval used gas-only equilibrium abundances rather than the default condensation equilibrium abundances, while the stellar spectrum was set to a blackbody in both codes in order to ensure that any differences in the results were due to differences in the planet model and not the stellar model.  We utilized nested sampling with 1000 live points for all retrievals.  For \platon, the package we used was \texttt{dynesty} \citep{speagle_2019}; for TauREx 3, it was \texttt{nestle}.

\begin{figure*}[ht]
  \centering
  \includegraphics
    [width=\textwidth]{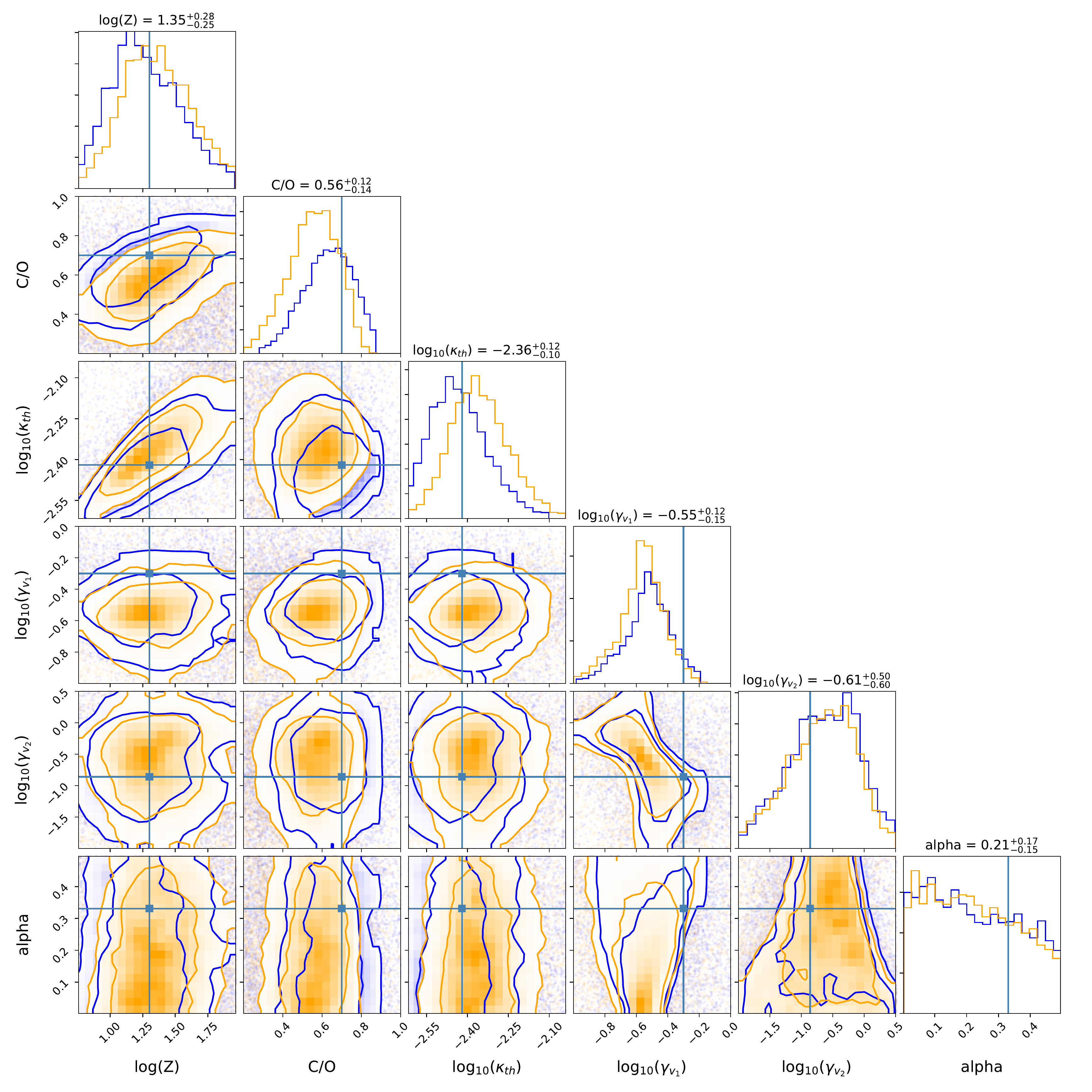}
  \caption{Posterior distributions retrieved by \platon (orange) and TauREx (blue) when both codes use the same opacities and include the same list of molecules.  The contours contain 68\% and 95\% of posterior mass.  The numbers on top of each column show the values inferred by \platon.  The teal horizontal and vertical lines show the truth values.}
\label{fig:platon_vs_taurex_corner}
\end{figure*}

We ran two comparisons.  In the first comparison, we generated the emission spectrum using \platon R=10,000 opacities, and included the same set of molecular opacities in both the \platon and TauREx retrievals: NH$_3$, CH$_4$, CO, CO$_2$, and H$_2$O.  In \platon, the opacities of all other molecules were set to 0.  Figure \ref{fig:platon_vs_taurex_corner} shows the results of this first retrieval comparison.  In general, the two codes give very similar posteriors.  The 1D posteriors of $\log_{\gamma_{v_1}}$, $\log_{\gamma_{v_2}}$, and $\alpha$ are indistinguishable.  The log(Z) and C/O posteriors show discrepancies at the 0.7$\sigma$ level because the equilibrium abundances of active gases differ by a few tens of percent between \platon and TauREx, which in turn is because the former includes many times more atoms and molecules in its calculations than the latter (see Subsection \ref{subsec:validation}).  As a result of these differences, \platon prefers slightly higher temperatures at a given pressure ($\Delta T \sim$30 K at 100 mbar), which is reflected in the slightly higher $\kappa_{th}$.

\begin{figure*}[ht]
  \centering
  \includegraphics
    [width=\textwidth]{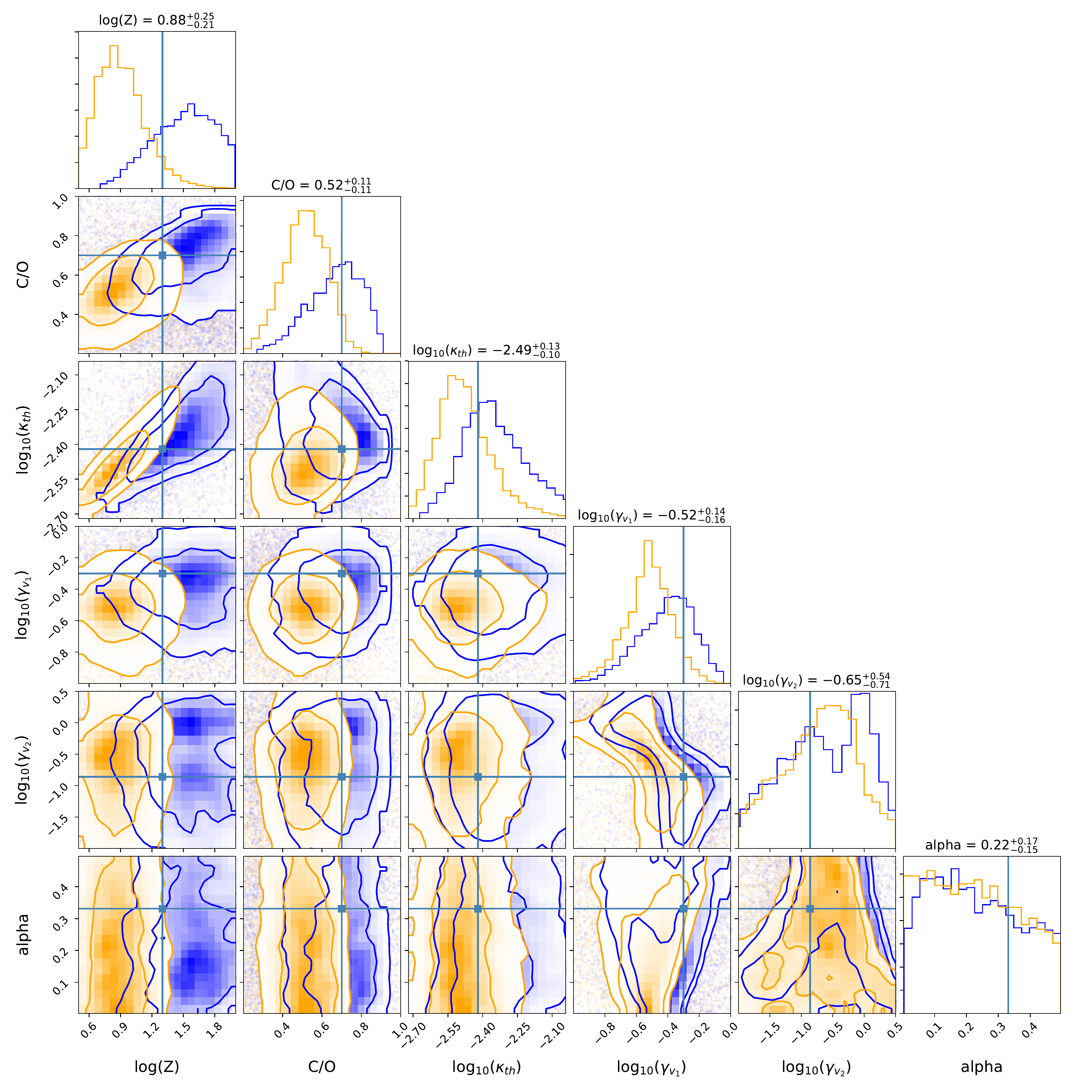}
  \caption{Same as Figure \ref{fig:platon_vs_taurex_corner}, but for the second comparision.  Here, TauREx opacities are used to generate the spectrum and for TauREx's retrieval.  The \platon retrieval uses its own opacities, with all molecules included.  The discrepancy is due to the inclusion of H$_2$S in \platon, but not in TauREx.}
\label{fig:platon_vs_taurex_corner_for_taurex_op}
\end{figure*}

In the second comparison, we generated the emission spectrum using TauREx R=15,000 opacities and used those same opacities in the TauREx retrieval.  For the \platon retrieval, we used the the full list of molecules with the same R=10,000 opacities as in the previous comparison.  The differences between the two retrievals therefore reflect both the effect of including different numbers of molecules, and differences in the line lists used for those molecules.  Figure \ref{fig:platon_vs_taurex_corner_for_taurex_op} shows the results of this second retrieval comparison.  Even though the \platon 1D posteriors are still consistent with the input planet parameters at the 1.7$\sigma$ level, they are more discrepant than in Figure \ref{fig:platon_vs_taurex_corner_for_taurex_op}, with \platon obtaining a metallicity 4$\times$ lower and a C/O ratio 0.18 lower than the TauREx retrieval.

Having obtained this result, the natural question to ask is what causes the discrepancy: \platon's newer line lists, or its inclusion of more molecules?  The answer is the latter.  We examined \platon's best fit model and found that it underestimated the planetary emission around 3.8 \um, where an opacity window caused a spike in planetary emission.  Removing molecules one by one from the atmosphere, we find that H$_2$S is the cause of the discrepancy: removing it alone from the atmosphere makes the best fit spectrum line up perfectly with TauREx's simulated data.  Indeed, when we repeat the \platon retrieval while including only the molecules that TauREx includes, the resulting posteriors are almost identical to those of Figure \ref{fig:platon_vs_taurex_corner}.  This underscores the importance of erring on the side of caution when choosing which active gases to include in a model.  Emission spectroscopy has the inconvenient property that it is the \textit{lack} of absorption that causes the most easily detectable emission--and so even a trace gas with a relatively low opacity can have a significant impact on the spectrum at wavelengths where other gases also have less absorption.

\section{Opacity update}
\label{sec:opacity}
One fundamental building block of any atmospheric code is the calculation of opacities.  There are three types of opacities we consider: scattering, line, and collisional.  As discussed in \cite{zhang_2019}, scattering opacities are calculated by \platon itself, and collisional opacities are calculated using the limited data available from HITRAN \textbf{\citep{richard_2012,karman_2019}}.  Line opacities are calculated from lists of transitions from one quantum state to another, giving the position, intensity, and broadening parameters of the transitions.  For this update to \platon we focused on line opacities.

In the original \platon release, our opacity data were taken directly from Exo-Transmit \citep{kempton_2017}.  Exo-Transmit, in turn, calculated its opacities from line lists generated from a large number of sources, listed in Table 2 of \cite{luppu_2014}.  These include HITRAN, HITEMP, private communications, \cite{freedman_2008}, and \cite{freedman_2014}, among many others.  Many of these line lists are outdated, proprietary, or both.  In addition, the program used to generate opacity data from the line lists is not public, making it difficult to reproduce our opacity calculations.

\begin{table}[ht]
  \centering
  \caption{Sources of line lists}
  \begin{tabular}{|p{2cm}|p{5.5cm}|}
  \hline
  	Source & Molecules\\
      \hline
    ExoMol & C$_2$H$_4$, CO, H$_2$CO, H$_2$S, H$_2$O, HCl, HCN, MgH, NH$_3$, NO, OH, PH$_3$, SH, SiH, SiO, SO$_2$, TiO, VO\\
    HITRAN 2016 & C$_2$H$_2$, C$_2$H$_6$, HF, N2, NO$_2$, O$_2$, O$_3$, OCS\\
    CDSD-4000 & CO$_2$\\
    Rey et al 2017 & CH$_4$\\
    NIST & Na, K\\
      \hline
  \end{tabular}
  \label{table:line_list_sources}
\end{table}

We address these shortcomings by regenerating \platon's opacity data using the public line lists in Table \ref{table:line_list_sources}.  For each molecule, we generate absorption cross sections from line lists using the method outlined in ExoCross \citep{yurchenko_2018}.  The cross sections are generated for 30 temperatures ($100-3000$ K in 100 K increments), 13 pressures ($10^{-4}- 10^8$ Pa in decade increments), and 4616 wavelengths ($0.3-30$ \um, with uniform spacing in logarithmic space).  The resolution of our wavelength grid is not high enough to resolve individual lines at typical atmospheric pressures (P $<$ 1 bar), leading to spikiness in the wavelength-dependent cross sections, and therefore in the final transit and secondary eclipse depths.  This approach to radiative transfer is called ``opacity sampling''.  The idea behind opacity sampling is that even though the sampling resolution is much lower than that needed to resolve individual lines, it is still much higher than the instrumental resolution, and the spikiness in the resulting models can be smoothed out by binning to instrumental resolution.

We generate cross sections by assuming a Voigt profile for every line, with the Gaussian standard deviation set by the temperature of the gas, and the Lorentzian portion set by the pressure broadening coefficients $\gamma_{\rm ref}$ and $n$:

\begin{align}
    \gamma(P,T) = \gamma_{\rm ref}\frac{P}{P_{ref}}\bigg(\frac{T_{\rm ref}}{T}\bigg)^n
\end{align}
where $\gamma_{\rm ref}$ and $n$ are expected to vary depending on the line considered and the species responsible for the broadening.  

\subsection{ExoMol}
ExoMol \citep{tennyson_2018} is a database of molecular line lists intended for modeling the atmospheres of exoplanets and cool stars.  The lists are generated using a combination of ab initio calculations and empirical data.  Many of the line lists represent significant improvements in completeness over the previous state of the art.  For example, POKAZATEL, the water line list, has 6 billion transitions.  This is an order of magnitude more than previous lists, and covers every possible transition between states below the dissociation energy of water \citep{polyansky_2018}.  Compared to the \cite{freedman_2008} line lists used by previous versions of \platon, the ExoMol line list has many times the number of transitions for water (6 billion vs. 200 million), NH$_3$ (10 billion vs. 34,000), H$_2$S (115 million vs. 188,000), PH$_3$ (50 billion vs. 20,000) and VO (377 million vs. 3.1 million).

The specific line lists we used are listed in Table \ref{table:exomol_line_lists}, together with the number of transitions they contain, their maximum temperature of validity, and citations to the associated papers.  For some molecules, ExoMol provides line lists for multiple isotopologues; in those cases we only include the most common isotopologue.  The exception to this rule is TiO, an important molecule where multiple isotopologues have comparable abundances.  For this molecule, we compute the absorption due to each isotopologue and add them in proportion to each isotopologue's abundance.

Although these ExoMol line lists are an improvement over what was available before, many are still incomplete.  The calculations only include states below a certain $J$ quantum number, and hence miss transitions between higher-energy states that become important at higher temperatures.  Thus, many line lists are not valid for the full range of temperatures supported by \platon, which ranges up to 3000 K.  For example, the C$_2$H$_4$ line list, despite having 50 billion lines, is only valid below 700 K.  For these molecules, we still generated cross sections for all temperatures.  The cross sections are likely to be under-estimated at high temperatures due to missing lines, but as of this writing, there is no better alternative.

ExoMol reports line broadening coefficients for hydrogen and helium whenever available.  In practice, however, there are no calculations or experimental data available for any broadening agent for the vast majority of lines.  For example, ExoMol reports no broadening coefficients at all for H$_2$S, CO, MgH, NO, OH, SiH, SiO, or VO.  For NO, it only reports air broadening coefficients, which we adopt for lack of a better alternative.  For the other molecules, ExoMol reports broadening coefficients for H$_2$ and He in a consistent format \citep{barton_2017}, relying on the handful of studies that have reported coefficients for a small number of lines while resorting to default values for the rest.  We used ExoMol broadening data when available and assumed that the broadening agent is a mixture of 85\% H$_2$ and 15\% He.  For all molecules where ExoMol broadening data is not available, we assumed that $\gamma_{ref} = 0.07$ and $n$=0.5 at a $T_{\rm ref} = 296 K$ and $P_{ref}$ = 1 bar.  $n$=0.5 is the theoretically expected value from classical calculations, while $\gamma_{ref} = 0.07$ is a typical value adopted by ExoMol as the default.  The only exception is C$_2$H$_4$, where we used broadening parameters measured by \cite{bouanich_2003} (H$_2$) and \cite{reuter_1993} (He) for 34 and 3 lines respectively by generating ExoMol-formatted broadening files from the measurements. 

\subsection{HITRAN 2016}
HITRAN \citep{gordon_2017} is a database of line lists sourced from a combination of observations, theory, and semi-empirical calculations.  It is intended for use at terrestrial temperatures, and has a line intensity cutoff that makes it inaccurate for higher temperatures.  Nevertheless, HITRAN is a valuable resource because it is the only source of line lists for many molecules.  HITRAN specifies line broadening parameters by including $\gamma$ and $n$ in the description of every line.  Although $\gamma_{H2}$ and $n_{H2}$ are included in HITRAN, very few lines have hydrogen broadening data.  Therefore, we chose $\gamma_{air}$ and $n_{air}$ as the broadening parameters for every line.

For most molecules not included in ExoMol, Exo-Transmit used (and \platon inherited) absorption data from HITRAN 2008.  We regenerate the absorption data using HITRAN 2016 \citep{gordon_2017}, which has expanded wavelength coverage and improved accuracy.  This update also fixed some errors in the old data that resulted from incorrect generation of absorption data from line lists.  The HAPI API makes it easy to retrieve line lists for all isotopologues at once, with intensities appropriately scaled to the isotopologue abundance.  Therefore, we considered all isotopologues for the molecules we took from HITRAN.

\subsection{CDSD-4000}
The Carbon Dioxide Spectroscopic Databank 4000 (CDSD-4000) is a line list meant for high temperatures, provided in a format similar to HITRAN.  It has significantly more lines than HITRAN or HITEMP, and to our knowledge it is the most complete publicly available line list for carbon dioxide.  CDSD-4000  has pressure broadening coefficients for air, which we adopt due to the absence of broadening coefficients for hydrogen or helium.

\subsection{Rey et al 2017}
The line list presented by \cite{rey_2017}, which we name Rey for convenience, is the first theoretical methane line list suitable for high temperature applications.  It is complete in the infrared range (0--13,400 cm$^{-1}$) up to a temperature of 3000 K, whereas the ExoMol line list ``10to10'' is only accurate to 1500 K \citep{yurchenko_2014}.  Rey also claims to be the first theoretical methane line list with line positions accurate enough for high resolution cross-correlation studies.  We have confirmed this claim by cross correlating brown dwarf models generated using both methane line lists to observational high-resolution spectra of a T brown dwarf, with all other opacity sources excluded.  The cross correlation peak is 4.5$\sigma$ with ``10to10'', but 15.3$\sigma$ with Rey, indicating far superior line positions.  In addition, with 150 billion lines, this line list is far more complete than either ExoMol's ``10to10'' line list (10 billion transitions) or the \cite{freedman_2008} line list that we used previously (200 million transitions).

\subsection{Voigt cutoff}
When generating absorption cross sections from line lists, the cutoff--namely, how far away from the line center the line is considered to end-- is an important source of error.  One could in principle omit the cutoff, but computational speed would suffer greatly.  In addition, omitting the cutoff does not necessarily lead to better results, as the Voigt profile is only an approximation to the true line profile \citep{ngo_2012} and is not accurate more than several Voigt widths away from the center.  However, truncating the lines too soon would result in an underestimate of the true opacity due to the omission of millions of line wings.

We adopted a cutoff of 25 cm\textsuperscript{-1} for all molecules for pressures less than or equal to 1 bar.  For pressures of 10 and 100 bar, we adopted a cutoff of 100 cm\textsuperscript{-1}.  For a pressure of 1000 bar, we adopted a cutoff of 1000 cm\textsuperscript{-1}.  This prescription was inspired by \cite{sharp_2007}, who adopt a cutoff of min(25$P$, 100) cm\textsuperscript{-1} where P is in bars.  \cite{hedges_2016} studied the effect of different cutoffs and concluded that the Sharp \& Burrows prescription significantly underestimates absorption at low pressures (P $<$ 0.01 bar), but is accurate from 0.01 to 100 bar.  We therefore modified the prescription to use 25 cm\textsuperscript{-1} for all pressures below or equal to 1 bar.  Pressure broadening coefficients are almost never measured or calculated at very high pressures ($100-1000$ bar), so our opacity data in this regime should be regarded as highly speculative.

\section{Other improvements}
\label{sec:improvements}
Aside from the opacity update and the eclipse depth calculator, many improvements have been made to \platon since the publication of the last paper.  These improvements introduce features, fix bugs, increase speed, and improve usability.  A comprehensive list can be found in our release notes, but we list a few of the most noteworthy updates below.

\begin{itemize}
  \item We now include H- opacity, calculated using the algorithm in \cite{john_1988}.  Because H- opacity is insignificant for most planets, we disable it by default.
  \item Nested sampling is now done by \texttt{dynesty} \citep{speagle_2019} rather than \texttt{nestle}.  Among other improvements, \texttt{dynesty} prints out the number of likelihood evaluations, log evidence (logz), and remaining evidence (dlogz) after each iteration.  dlogz is an indicator of how far the algorithm is from completion.  \texttt{nestle} did not have this indicator, which made waiting a frustrating experience.
  \item The eclipse depth calculator now evaluates the single integral in Equation \ref{eq:emergent_flux_E2} instead of the double integral in Equation \ref{eq:emergent_flux}, making it much faster, many times more memory efficient, and more accurate.
  \item Data arrays are transposed so that the wavelength index increases the fastest, followed by the pressure index, followed by the temperature index.  This improves cache locality, which speeds up the code by a factor of 1/3.
  \item The number of atmospheric layers is decreased from 500 to 250, improving speed by 40\% while increasing numerical error by only $\sim$1\%.  In addition, we now use improved interpolation methods to further decrease numerical errors.
\end{itemize}

\subsection{High resolution opacities}
\platon has a clean separation between data and code.  As a result, all that is needed to operate \platon at an arbitrary wavelength range and resolution is the appropriate opacity data files. No code changes are required. Since we published \cite{zhang_2019}, we have generated high-resolution opacity data files for a variety of applications.  This includes studying the atmospheres of cold brown dwarfs, of the ultra-hot super Earth 55 Cnc e, and of HD 189733b (see Subsection \ref{subsec:hd189_high_res}). 

We now make these opacities public to enable anyone to perform line-by-line calculations with \platon\footnote{\url{https://www.astro.caltech.edu/platon}}.  All opacities have a resolution of R=375,000, and are calculated at the wavelengths indicated by wavelengths.npy.  These opacities can be used by deleting all files from the ``Absorption" folder in \platon's data directory, putting the downloaded absorb\_coeffs files into the directory, and replacing the wavelengths.npy file in the data directory with the one in the downloaded zip file.  The user must use a blackbody stellar spectrum (by passing stellar\_blackbody=True to compute\_depths) when using high resolution opacities.  The user can also generate their own high-resolution opacity files using publicly available codes such as ExoCross \citep{yurchenko_2018} or HELIOS-K \citep{grimm_2015}.  As long as they are in the same format as the \platon data files, \platon will accept them with no code changes.

\begin{table}[ht]
  \centering
  \caption{High resolution absorption data}
  \begin{tabular}{|p{1.1cm}|p{1.8cm}|p{4.5cm}|}
  \hline
  	Filename & Wavelengths (\um) & Molecules\\
      \hline
    hispec & $0.94-2.43$ & CH$_4$, CO, H$_2$O, H$_2$S, HCl, HCN, MgH, NH$_3$, NO$_2$, NO, O$_2$, O$_3$, OH, SH, SiH, SiO, SO$_2$, TiO, VO\\
    $Y$\_band & $1.020-1.086$ & CH$_4$, H$_2$O, NH$_3$\\
    $K$\_band & $1.89-2.40$ & C$_2$H$_2$, CH$_4$, CO, H$_2$O, HCN, Na, NH$_3$, SiO\\
    $L$\_band & $2.86-3.70$ & C$_2$H$_2$, CH$_4$, CO, H$_2$O, H$_2$S, HCN, Na, SiO\\
    $L$\_band2 & $3.51-4.08$ & CH$_4$, CO$_2$, H$_2$O, HCN\\
      \hline
  \end{tabular}
  \label{table:high_res_lists}
\end{table}

We note that in principle other codes, such as TauREx, can also be used to calculate high resolution spectra if they are provided with custom user-generated high resolution opacity files. To the best of our knowledge, however, \platon is the only code to make such opacity files publicly available.

\subsection{Correlated-$k$}
The gold standard of radiative transfer is the line-by-line method \citep{marley_2015}: calculating the transit or eclipse depth on a wavelength grid fine enough to resolve individual molecular lines, which typically requires $R >> c/\sqrt{\frac{kT}{\mu m_H}} \approx 200,000$.  The results are then binned to instrumental resolution.  For many applications, including \platon, this is computationally prohibitive.  Opacity sampling sacrifices accuracy for speed by performing radiative transfer at a much lower resolution--R=1000, in the case of \platon--and binning the depths thus obtained to instrumental resolution.  In order for the survey to approximate the results of the census, a large sample size is required, meaning opacity sampling at R=1000 is only accurate if the user is binning to resolutions much below 1000.  To take a concrete example, suppose \platon is used to calculate the eclipse depth of HD 189733b over the 1.40--1.42 \um band using the R=10,000 opacities.  \platon would sample 140 wavelengths within this band and calculate the eclipse depth at each wavelength.  These 140 eclipse depths would have a mean of 53 ppm and a standard deviation of 25 ppm.  Therefore, the error caused by opacity sampling is $25/\sqrt{140}$=2 ppm.  If the R=1000 opacities were used instead, this error would be $25/\sqrt{14}=7$ ppm.  Fortunately, 7 ppm is several times lower than the error of the WFC3 observations, but this will not be the case for every combination of planet, instrument, and wavelength band.

The correlated-$k$ method \citep{lacis_oinas_1991} improves upon opacity sampling by taking into account the distribution of opacities within the passband.  For example, it calculates the 10th percentile eclipse depth by using the pre-calculated 10th percentile of all molecular opacities, and likewise for the 20th percentile, 30th percentile, etc.  Since the eclipse depth varies smoothly with percentile, we can use numerical integration to find the average eclipse depth.  In effect, the correlated-$k$ method converts an integration over wavelength into an integration over percentile; it replaces the integration of a highly non-smooth function $\int_{\lambda_1}^{\lambda_2}{f(\lambda)d\lambda}$ with the integration of a smooth function $\int_0^1{f'(g)dg}$, which is in turn evaluated by Gaussian quadrature ($\sum_{i=1}^N {w_i f'(g_i)}$), where g is the percentile divided by 100.  $f(\lambda)$ can be $R_p(\lambda)$, the emergent flux $F_p(\lambda)$, or any other radiative quantity, so long as its only dependence on wavelength is through the opacity, and so long as it varies smoothly with opacity.

Earlier versions of \platon performed radiative transfer via opacity sampling at a default resolution of 1000 with an optional R=10,000 mode.  We now give the user the option to choose between correlated-$k$ (R=100) and opacity sampling.  The correlated-$k$ method provides the accuracy of R=50,000 opacity sampling for typical exoplanet applications, but runs at the same speed as the old default R=1000 \platon mode. Our implementation of correlated-$k$ rests on two approximations:

\begin{enumerate}
    \item At any given wavelength $\lambda$ within the band, $g(\lambda)$ is the same for every layer.  That is, if a layer is more opaque at a certain wavelength than at x\% of other wavelengths within the band, all other layers must also be more opaque at that wavelength than at x\% of other wavelengths.  
    \item For each layer, if the individual gases were to be separated out, $g(\lambda)$ would be equal for all gases with significant opacity.  That is, if one molecule absorbs more strongly at a certain wavelength than at x\% of other wavelengths within the band, the same must be true for all other molecules.
\end{enumerate}

The first assumption is the defining assumption of the correlated-$k$ method, and explains its name: the opacity $\kappa$ is assumed to be correlated throughout the atmosphere under consideration.  For an atmosphere with one species with exactly one absorption or emission line within the band, it is exactly true.  For an atmosphere with two layers, each of which is exclusively composed of a different gas, it is very inaccurate.  A real atmosphere is in between these two extremes: the region dominating the features in a transmission or emission spectrum typically spans $\sim$200 K in temperature (Figure \ref{fig:contrib_and_TP}) and 1-2 orders of magnitude in pressure (see Figure \ref{fig:contrib_and_TP} and \ref{fig:transit_contrib}), which generally means that there is no change in the dominant gas absorber. 

The second assumption, however, is only true when one molecule dominates the opacity.  If two molecules contribute equally to the opacity, the assumption is no longer valid, as the absorption lines of different molecules will not in general overlap.  In fact, the opposite assumption is more accurate: namely, that the opacities are completely uncorrelated between different gases.  \cite{lacis_oinas_1991} take this approach, but adopting this assumption naively for \platon would require O($n^N$) radiative transfer calculations, where $n$ is the number of discrete $g$ values adopted (for us, 10) and $N$ is the number of gases (for us, 30), making these computations intractable.  There are methods of merging the opacity distributions of multiple gases that do not scale exponentially--including the `random overlap with resorting and rebinning' method introduced by \cite{lacis_oinas_1991} and named by \cite{amudsen_2017}.  These are more complicated to implement, and we may incorporate them into a future release of \platon.  The partially correlated approach attempts to take into account the correlations between gases \citep{zhang_2003}, but these sophisticated schemes are beyond the scope of \platon.

We note that the simplicity of our approach comes at a cost: it systematically overestimates transmittance in most cases \citep{zhang_2003}.  This overestimation is easy to understand with a toy scenario.  Consider a gas with binary absorption properties: at 50\% of wavelengths it has infinite absorption, while at the other wavelengths it has zero absorption.  This gas would have a transmittance of 50\%.  Now consider adding a second gas, also with binary absorption properties.  If the two gases have perfectly correlated absorption (which is our assumption), their absorption peaks fall on top of each other, and transmittance is still 50\%.  If their absorption is not perfectly correlated, the absorption peaks of the second gas block some of the light that would have went through the first gas, and total transmittance is less than 50\%.  If their absorption is perfectly anti-correlated, the total transmittance would be 0\%.

In practice, \platon uses the following correlated-$k$ algorithm:

\begin{enumerate}
    \item (Pre-computed) Compute the absorption coefficients of each atom/molecule at each temperature, pressure, and wavelength grid point, with a spectral resolution of R=50,000.  Correlated-$k$ coefficients are generated from the absorption coefficients with a resolution of R=100.
    \item Divide the wavelength range under consideration into bands, with each band having a width of $\lambda/100$
    \item For each band, compute the transit/eclipse depth at 10 different opacity percentiles, and combine them via Gaussian quadrature.  The transit depth at the 16th opacity percentile (for example) is calculated by assuming every gas, at every temperature and pressure, has an absorption coefficient equal to the pre-calculated 16th percentile absorption coefficient for that band at that temperature and pressure.  We use 10 Gaussian quadrature points, which is sufficient to keep the integration error below 1\% in most cases \citep{lacis_oinas_1991,goody_1989}.
    \item The R=100 transit or eclipse depths are then binned to the user-specified wavelength bins using the methods described in our first paper \citep{zhang_2019}.
    \label{list:correlated_k_assumptions}
\end{enumerate}

We performed an experiment to deduce the accuracy of the correlated-$k$ algorithm compared to a line-by-line calculation. The transit spectrum of a hot Jupiter (modelled after HD 209458b) was computed from $0.95-2.4$ \um using two methods: a line-by-line calculation at R=375,000 binned to R=100, and a correlated-$k$ calculation using R=100 opacities with 10 Gaussian quadrature points.  For the nominal model, which is dominated by water opacity, correlated-$k$ performs extremely well and has a maximum error of only 8 ppm.  For a model engineered to include three molecules with comparably significant absorption correlated-$k$ still performs well, with a maximum error of 300 ppm.  The emission spectrum tells a similar story.  Correlated-$k$ is accurate to 0.1\% for the nominal model, and to 3.5\% for the pessimistic model.

These tests also demonstrate that although correlated-$k$ is very accurate, its errors are not random.  Correlated-$k$ almost always underestimates the transit depths and overestimates the eclipse depths.  This is a consequence of overestimating the transmittance, which in turn is because (contrary to the second assumption above) the absorption properties of two molecules are in general not strongly correlated.

\subsection{Beta features}
Since we do not currently plan to write a third \platon paper, we include a list of beta features that will likely become part of the official \platon.  This list also serves to illustrate what is possible with minimal hacking.  All of these features were created as a result of requests from \platon users other than the authors.  Users are highly encouraged to contact the authors to suggest new features or improvements to existing features.

\platon does not calculate disequilibrium chemistry from first principles, nor does it compute self-consistent temperature-pressure profiles.  It is often useful to take abundance and temperature profiles from elsewhere and plug them into \platon, using it as a radiative transfer engine to predict transit and eclipse depths.  This is currently easy to do for vertically-constant abundances, but not for vertically-variable abundances.  To make the latter possible, we created the branch custom\_abundances on the GitHub repository.  Examples of how to use it are found in examples/plot\_transit\_custom\_abunds.py and examples/plot\_eclipse\_custom\_abunds.py.  

Some metallic species are not included in \platon by default, but may become important in the optical for ultra-hot Jupiters.  These include Ca, Fe, Ni, and Ti.  We make these opacities available at \url{https://www.astro.caltech.edu/platon/metal_opacities/}.  These can be used by placing them in PLATON\_DIR/data/Absorption and adding the atoms to PLATON\_DIR/data/species\_info.  These atoms are not incorporated into the equilibrium chemistry calculation, but the user can easily specify vertically constant abundances for them.  We describe the procedure in \platon's online documentation.

Lastly, Na and K each have two very strong lines in the optical, where the atmosphere is transparent enough that their far wings may become significant.  Unfortunately, while the lines cores are accurately described by a Voigt profile, the Voigt profile can underestimate far wing absorption by orders of magnitude \citep{allard_2016,allard_2019}.  More accurate line profiles for these atoms were recently published by \cite{allard_2016} (K) and \cite{allard_2019} (Na) using a semi-classical theory and assuming broadening by molecular hydrogen only.  We use these line profiles to generate \platon-friendly absorption coefficients at R=1000 and R=10,000, found at \url{https://www.astro.caltech.edu/platon/metal_opacities/}.  By overwriting PLATON\_DIR/data/Absorption/absorb\_coeffs\_Na.npy and PLATON\_DIR/data/Absorption/absorb\_coeffs\_K.npy with these coefficients, the user can generate much more accurate hydrogen-broadened alkali line profiles with \platon.

\section{Retrieval on HD189733\lowercase{b}}
\label{sec:hd189733b}
\subsection{Published data sets}
HD 189733b is one of the most favorable exoplanets for atmospheric characterization.  It is a transiting hot Jupiter orbiting an exceptionally close (20 pc) K star with a $H$ band magnitude of 5.6, and was one of the earliest transiting planets discovered \citep{bouchy_2005}.  To demonstrate \platon's new abilities, we perform a joint retrieval on the best available optical and near infrared transit and secondary eclipse data for HD 189733b from \emph{HST} and \emph{Spitzer}.  To our knowledge, this is the first joint transit and secondary eclipse retrieval for this planet in the literature, as well as the most comprehensive set of both transit and secondary eclipse data assembled for a retrieval to date.  The fixed stellar and planetary parameters are listed in Table \ref{table:real_planet_params}.  The transit depths we adopt are listed in Table \ref{table:adopted_transit_depths}, while the eclipse depths are listed in Table \ref{table:adopted_eclipse_depths}.  

HD 189733b has been observed in transmission with \emph{HST}/STIS \citep{sing_2011} and WFC3 \citep{gibson_2012,mccullough_2014}, \emph{Spitzer} in all five IRAC bands \citep{tinetti_2007,beaulieu_2008,agol_2010,desert_2011,morello_2014}, and \emph{Spitzer}/MIPS at 24 \um \citep{knutson_2009}.  \cite{pont_2013} carried out a uniform re-analysis of all transit data obtained to date including corrections for stellar activity; we utilize their transmission spectral data in our analysis.  This planet was also observed in transit by \emph{HST}/NICMOS in spectroscopic \citep{swain_2008} and photometric \citep{sing_2009} modes, but this instrument was less stable than WFC3, and the reliability of the spectroscopic NICMOS observations was questioned in a subsequent study \citep{gibson_2011,deming_seager_2017}.  We therefore exclude these older NICMOS observations from our analysis.  We also exclude the higher resolution observations of the sodium line published in \citep{huitson_2012}, as \platon is not designed to model absorption at the very low pressures probed by the core of this line.

In emission, HD 189733b has been observed with \emph{HST}/STIS \citep{evans_2013}, \emph{HST}/NICMOS \citep{swain_2009}, \emph{HST}/WFC3 \citep{crouzet_2014}, \emph{Spitzer}/IRAC in all four bands \citep{knutson_2007,charbonneau_2008,agol_2010,knutson_2012}, \emph{Spitzer}/IRS at $5-14$ \um \citep{deming_2006,grillmair_2008,todorv_2014}, and \emph{Spitzer}/MIPS at 24 \um \citep{knutson_2009}.  Because \platon does not model reflected light, we limit our retrieval to the infrared data only, which are expected to be dominated by thermal emission.  As with the transmission spectrum, we exclude the NICMOS observations from our retrieval.  We also exclude the Spitzer/IRS emission spectrum, as over the years different groups have obtained contradictory results. Most recently, \cite{todorv_2014} found that the overall amplitude of the IRS eclipse depth can shift up and down depending on the method used to correct for systematics.  Indeed, a comparison to our best fit model spectrum reveals that these data are consistent with the broadband observations if they are shifted upwards by 30\%.  In addition to transit and secondary eclipse observations, HD 189733b's phase curve has also been measured in the 3.6, 4.5, 8.0, and 24 \um \emph{Spitzer} bands \citep{knutson_2007,knutson_2009,knutson_2012}.  

Based on the observations listed above, HD 189733b is one of the most extensively observed transiting planets to date.  Previous studies of HD 189733b's optical transmission spectrum found that it appears to have a strong scattering slope and attenuated absorption features due to the presence of high-altitude scattering particles \citep{pont_2008,sing_2011,pont_2013}.  These scattering particles are possibly some form of silicate condensate \citep{lecavelier_2008,lee_2015,helling_2016}.  Because this planet is expected to be tidally locked, it should develop a super-rotating equatorial band of wind that transports heat from the day side to the night side (e.g., \citealt{showman_2011}).  It is observed to have a relatively modest day-night temperature gradient \citep{knutson_2007,knutson_2009,knutson_2012} and models predict that it may also have spatially inhomogeneous cloud coverage \citep{lee_2015,lines_2018}.

In the infrared the effect of the scattering particles on HD 189733b's transmission spectrum is reduced. \cite{mccullough_2014} report the detection of a spectroscopically resolved water feature at 1.4 \um that is consistent with the model of scattering aerosol reported by \cite{pont_2013}, although they also argue that the optical slope can be explained by stellar activity alone (see Section \ref{subsec:activity}).  In emission, previous studies have detected spectroscopically resolved water absorption at 1.4 \um \citep{crouzet_2014} and (debatably) in the mid-infrared \citep{grillmair_2008,todorv_2014}, and have placed additional constraints on the abundances of carbon monoxide, carbon dioxide, and methane based on the relative depths of the broadband \emph{Spitzer} secondary eclipse data (e.g., \citealt{line_2010,lee_2012}).  We discuss the results of these retrievals in more detail in Section \ref{sec:comparison}.

\subsection{\platon retrieval}
When modeling HD 189733b's transmission spectrum, our retrieval uses a complex refractive index with a real component of 1.7, a value in between that of MgSiO$_3$ and SiO$_2$ (which have $n \sim 1.5$ at optical wavelengths) and TiO$_2$ (which has $n \sim 2.4$).  Since the true composition of condensates in the atmosphere is unknown, and multiple condensates may well be important, we allow the imaginary component of the refractive index to vary as a free parameter in our fit.  We do not include clouds in our dayside models; even if the clouds observed on the terminator extended over the entire dayside, we would expect them to have lower optical depths when viewed in emission at infrared wavelengths (e.g., \citealt{fortney_2005}).  Indeed, previous retrieval studies of HD 189733b's dayside atmosphere have found that cloud-free models provide a good fit to the available data (e.g. \citealt{barstow_2014}).

\begin{figure*}[ht]
  \centering \subfigure {\includegraphics
    [width=\textwidth]{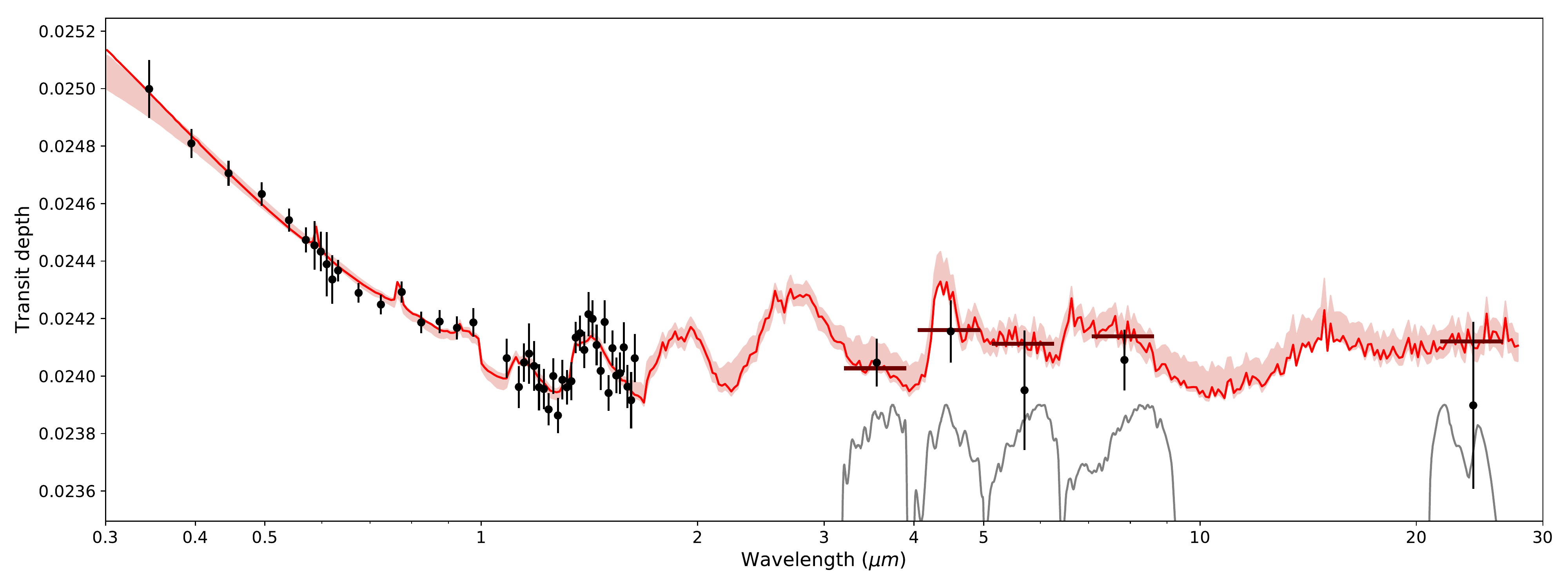}}\qquad
  \subfigure {\includegraphics
    [width=\textwidth]{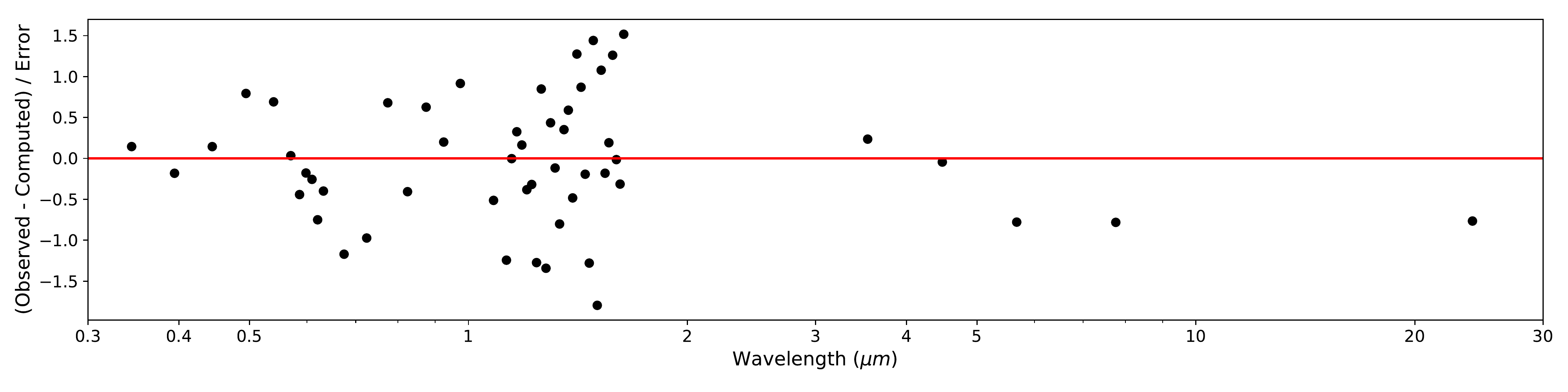}}
    \caption{\textbf{Top:} Best fit transit spectra from retrieval on HD 189733b.  The best-fit \platon model and corresponding $1\sigma$ uncertainty window are shown as a red line and red shaded region, respectively.  The activity-corrected observations drawn from \cite{pont_2013} and listed in Table \ref{table:adopted_transit_depths} are shown as black filled circles.  \textbf{Bottom:} difference between observed and computed transit depths, in units of measurement error $\sigma$.  Adopting the nightside emission pollution correction of \cite{kipping_2010} would reduce the errors on the 8\um and 24\um points by 0.6$\sigma$ and 0.4$\sigma$ respectively, bringing the observations into nearly perfect agreement with the model.}
\label{fig:best_fit_transits}
\end{figure*}

\begin{figure*}[ht]
  \centering \subfigure {\includegraphics
    [width=\textwidth]{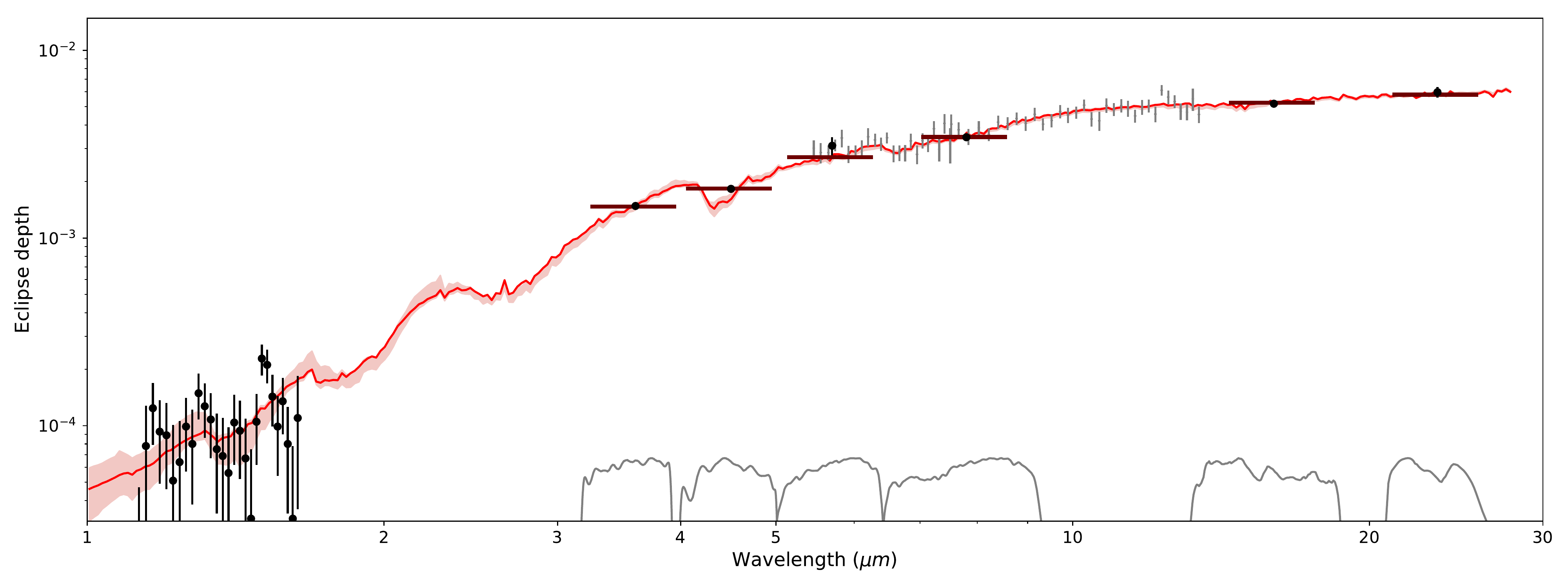}}\qquad
  \subfigure {\includegraphics
    [width=\textwidth]{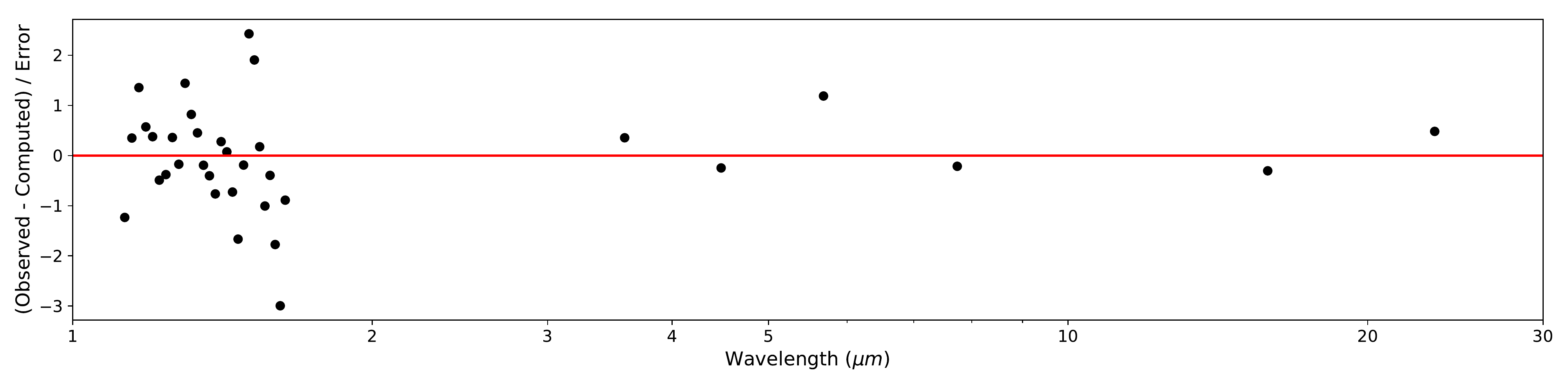}}
    \caption{\textbf{Top:} Best fit eclipse spectra from retrieval on HD 189733b.  The best-fit \platon model and corresonding $1\sigma$ uncertainty window are shown as a red line and red shaded region, respectively.  Data used in the fit (see list in Table \ref{table:adopted_eclipse_depths}) are shown as black filled circles. The IRS eclipse depths from \cite{todorv_2014} are shown in grey and were not included in the retrieval.  These eclipse depths have been shifted up by 30\% to match the model--a plausible shift, since \cite{todorv_2014} mentions that its results are 20\% below those of \cite{grillmair_2008}. \textbf{Bottom:} difference between observed and computed secondary eclipse depths, in units of measurement error $\sigma$.}
\label{fig:best_fit_eclipses}
\end{figure*}

We carry out our retrievals using the \texttt{dynesty} nested sampling package with static sampling and R=10,000 opacities.  We used 1000 live points, a convergence criteria of $\Delta$log(z) = 1, multi-ellipsoidal bounds, and a random-walk sampling method. The fixed parameters, listed in Table \ref{table:real_planet_params}, are the stellar radius, stellar temperature, and planetary mass.  The free parameters, listed in Table \ref{table:retrieved_parameters}, are the planetary radius $R_p$ at 1 bar, the metallicity $Z$ relative to solar, the C/O ratio, the limb temperature $T_{limb}$, the \cite{line_2013} T/P profile parameters (thermal opacity $\kappa_{th}$, visible-to-thermal opacity ratio of first visible stream $\gamma$, visible-to-thermal opacity ratio of second visible stream $\gamma_2$, percentage apportioned to the second visible stream $\alpha$, effective albedo $\beta$), mean haze particle radius $r_m$, haze particle number density $n$, ratio of haze scale height to gas scale height $h_{frac}$, WFC3 instrumental offsets ($\Delta_{wfc3,t}$ for transit and $\Delta_{wfc3,e}$ for eclipse), and the imaginary portion of the haze refractive index $k$.   In Figure \ref{fig:best_fit_transits} and \ref{fig:best_fit_eclipses}, we show the best fit transit and eclipse spectra from our retrieval.  Our best fit model is a good fit overall to the data, with a $\chi^2$ of 30.7 for the transit spectrum and 38.3 for the eclipse spectrum for a total $\chi^2=69$.  With 52 transit depths, 34 eclipse depths, and 15 free parameters, the p value is 0.55.  The single largest point of disagreement (3.1$\sigma$) between model and data occurs at the very red end of the WFC3 emission spectrum, where edge effects may impact the reliability of data.  

\begin{figure*}[ht]
  \centering 
  \includegraphics[width=\linewidth]{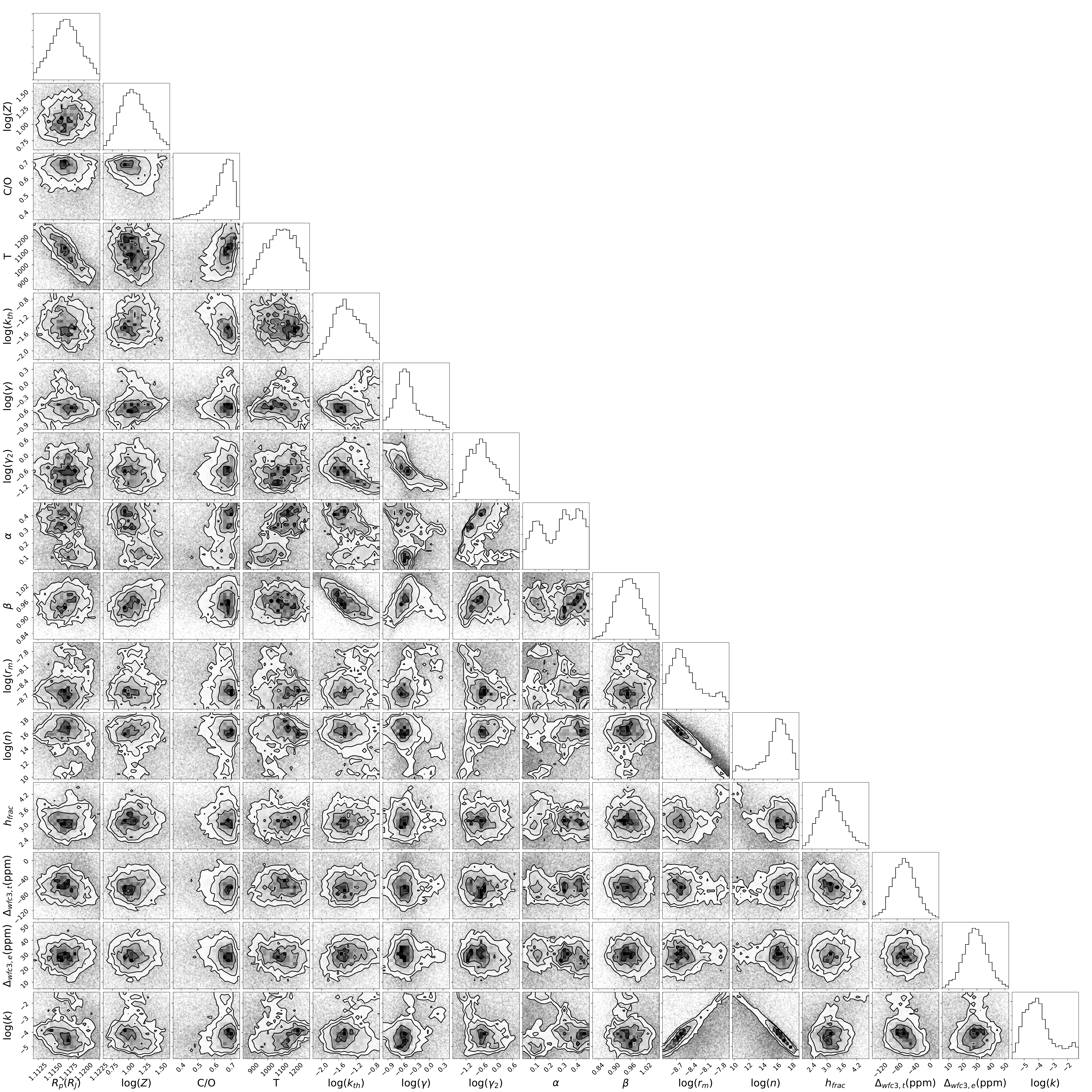}
  \caption{Posterior distribution of the fiducial retrieval on HD 189733b.}
\label{fig:big_corner}
\end{figure*}

\begin{table}[ht]
  \centering
  \caption{Retrieved parameters}
  \begin{tabular}{C C c C}
  \hline
  	\textrm{Parameter} & \textrm{Posterior} & \textrm{Best} & \textrm{Prior} \\
      \hline
      R_p\textrm{ (R$_\textrm{J}$)} & 1.117 \pm 0.002 & 1.114 & [1.11,1.13]\\
      \log_{10} Z & 1.08_{-0.20}^{+0.23} & 0.956 & [-1,3]\\
      C/O & 0.66_{-0.09}^{+0.05} & 0.69 & [0.2,2]\\
      T_{limb} \textrm{ (K)} & 1089_{-120}^{+110} & 1203 & [500, 1300]\\
       \log_{10} \kappa_{th} \textrm{ (m$^2$ kg$^{-1}$)} & -1.40_{-0.32}^{+0.40} & -1.44 & [-5,0]\\
      \log_{10} \gamma & -0.51_{-0.21}^{+0.36} & -0.66 & [-4,1]\\
      \log_{10} \gamma_2 & -0.58_{-0.50}^{+0.62} & -0.47 & [-4,1]\\
      \alpha^* & < 0.47 & 0.441 & [0,0.5]\\
      \beta & 0.95 \pm 0.05 & 0.938 & [0.5,2]\\
      \log_{10} r_m $\textrm{(m)}$^* & <-7.8 & -8.59 & [-9,-5]\\
      \log_{10} n (m^{-3}) & 16.0_{-2.7}^{+1.5} & 14.4 & [8,21]\\
      h_{frac} & 3.2_{-0.5}^{+0.6} & 3.19 & [0.5,5]\\
      \Delta_{wfc3,t} \textrm{(ppm)} & -63 \pm 28 & -72 & 0 \pm 100\\
      \Delta_{wfc3,e} \textrm{(ppm)} & 29 \pm 9 & 33 & 0 \pm 39\\
      \log_{10} k^* & *<-1.7 & -3.86 & [-6,0]\\
      \hline
  \end{tabular}
  \tablecomments{$^*$For these parameters, the 95th percentile upper bound is reported.}
  \tablecomments{The `best' column reports the parameters of the best fit model.}
  \label{table:retrieved_parameters}
\end{table}

\begin{figure*}[ht]
  \centering \subfigure {\includegraphics
    [width=0.5\textwidth]{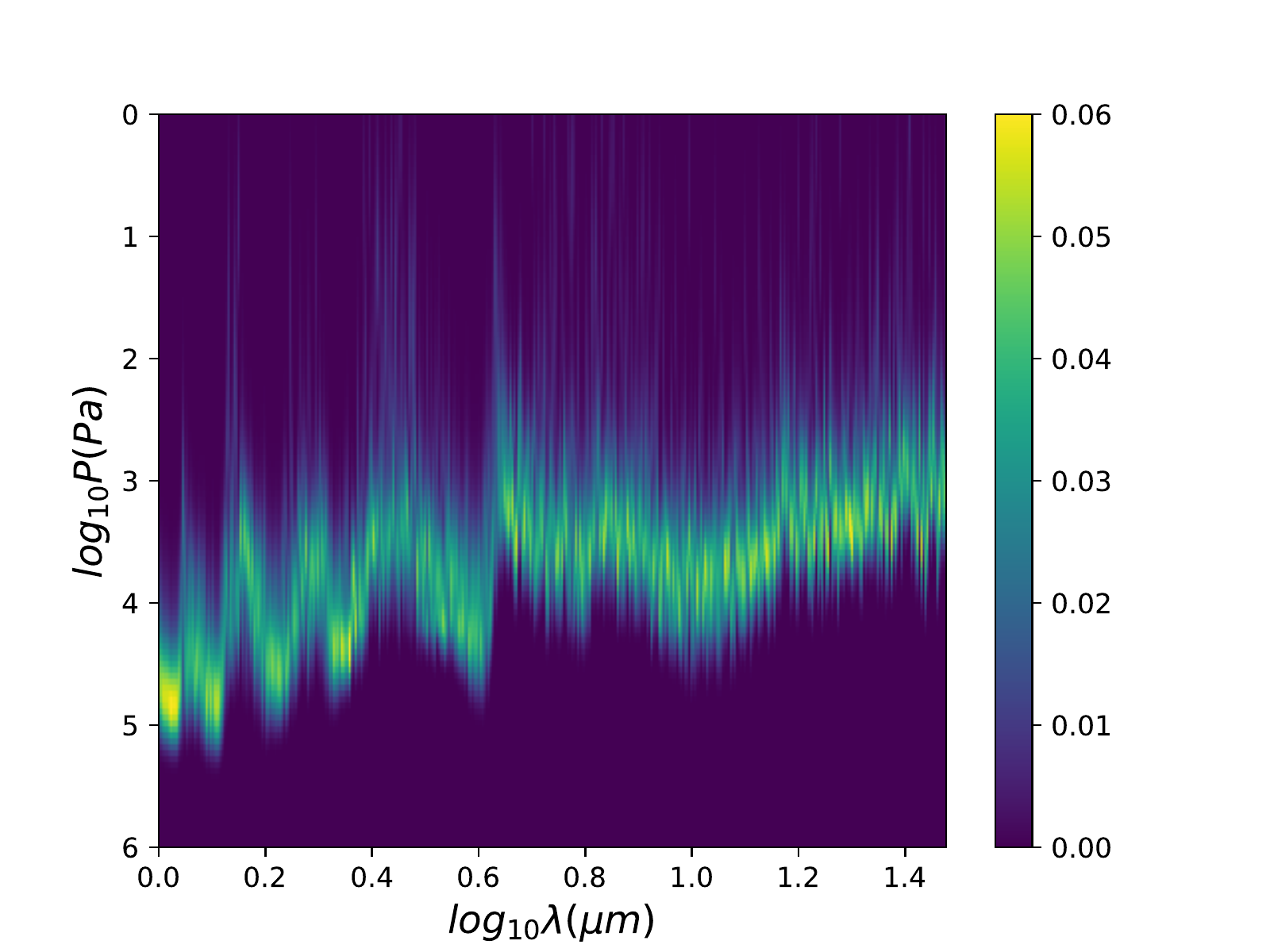}}\subfigure {\includegraphics
    [width=0.5\textwidth]{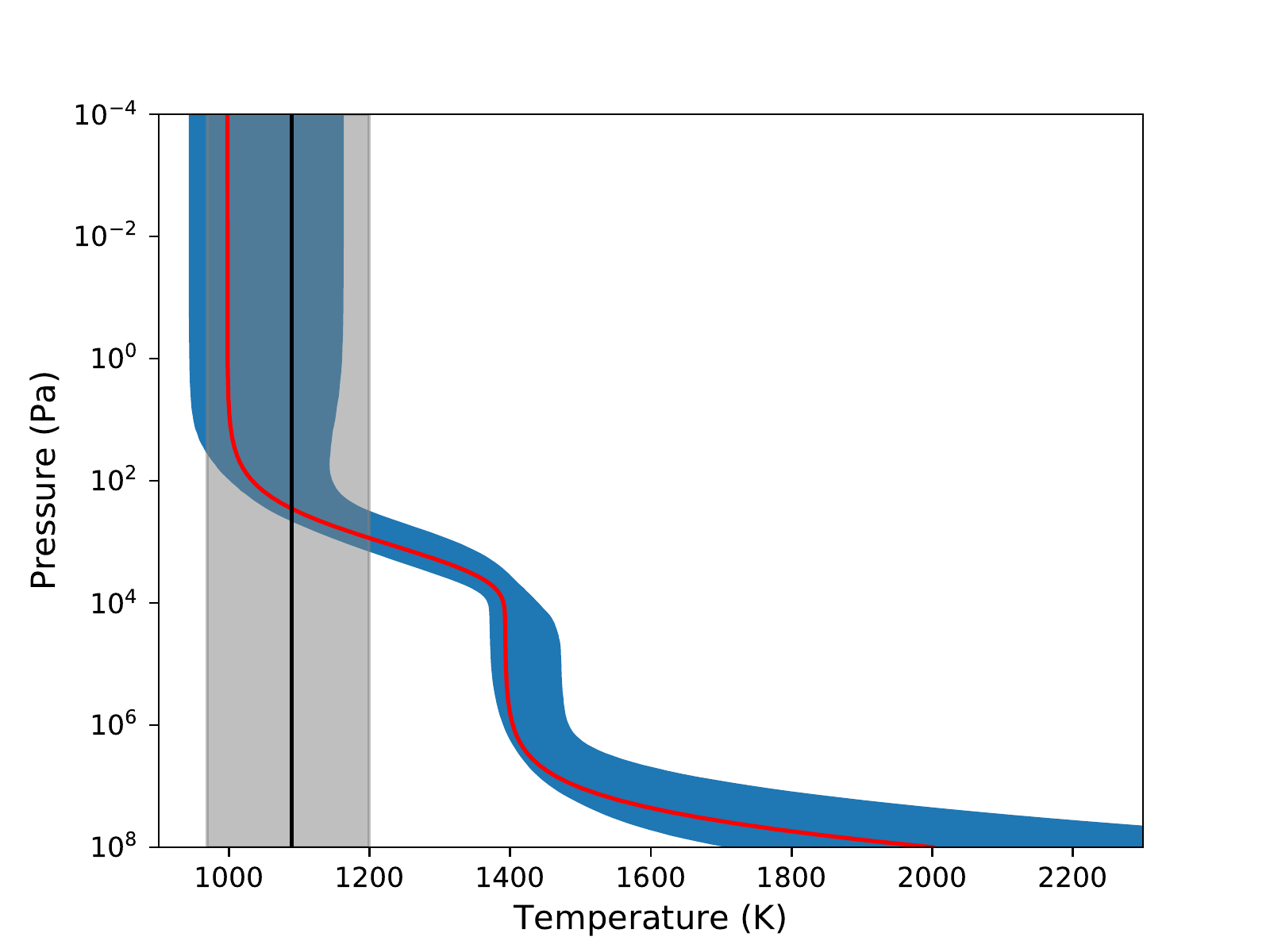}}
    \caption{Left: emission contribution function of HD 189733b, as indicated by the best fit solution.  Right: the T/P profile as indicated by the best fit solution (red), along with the 2$\sigma$ uncertainties on the T/P profile (blue).  The median limb temperature is indicated in black, while the 1$\sigma$ uncertainty in limb temperature is indicated in gray.}
\label{fig:contrib_and_TP}
\end{figure*}

\begin{figure}[ht]
  \centering 
  \includegraphics[width=\linewidth]{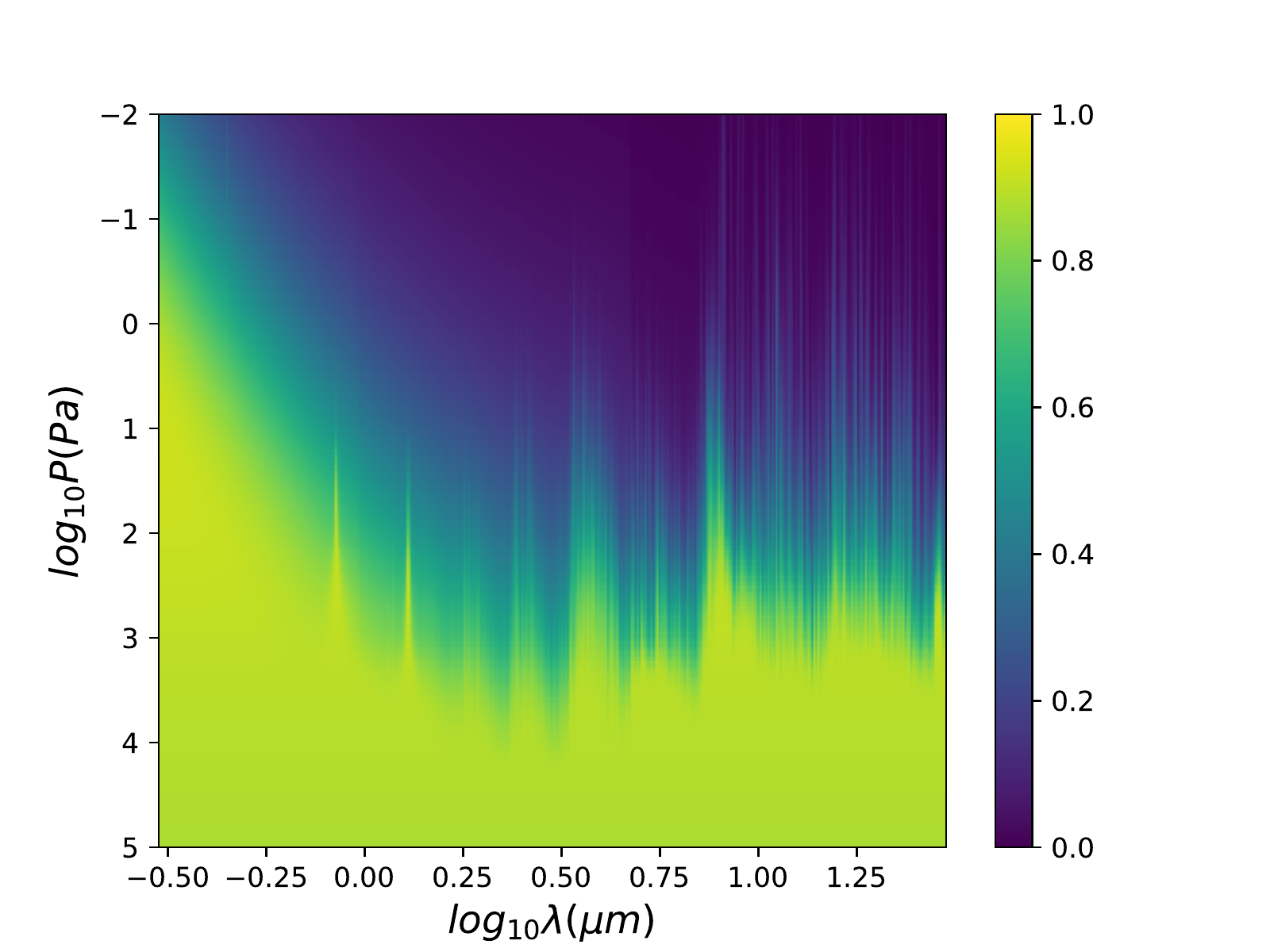}
  \caption{Contribution function of the transmission spectrum.  The absolute scale is in arbitrary units.}
\label{fig:transit_contrib}
\end{figure}

\begin{figure}[ht]
  \includegraphics
    [width=0.5\textwidth]{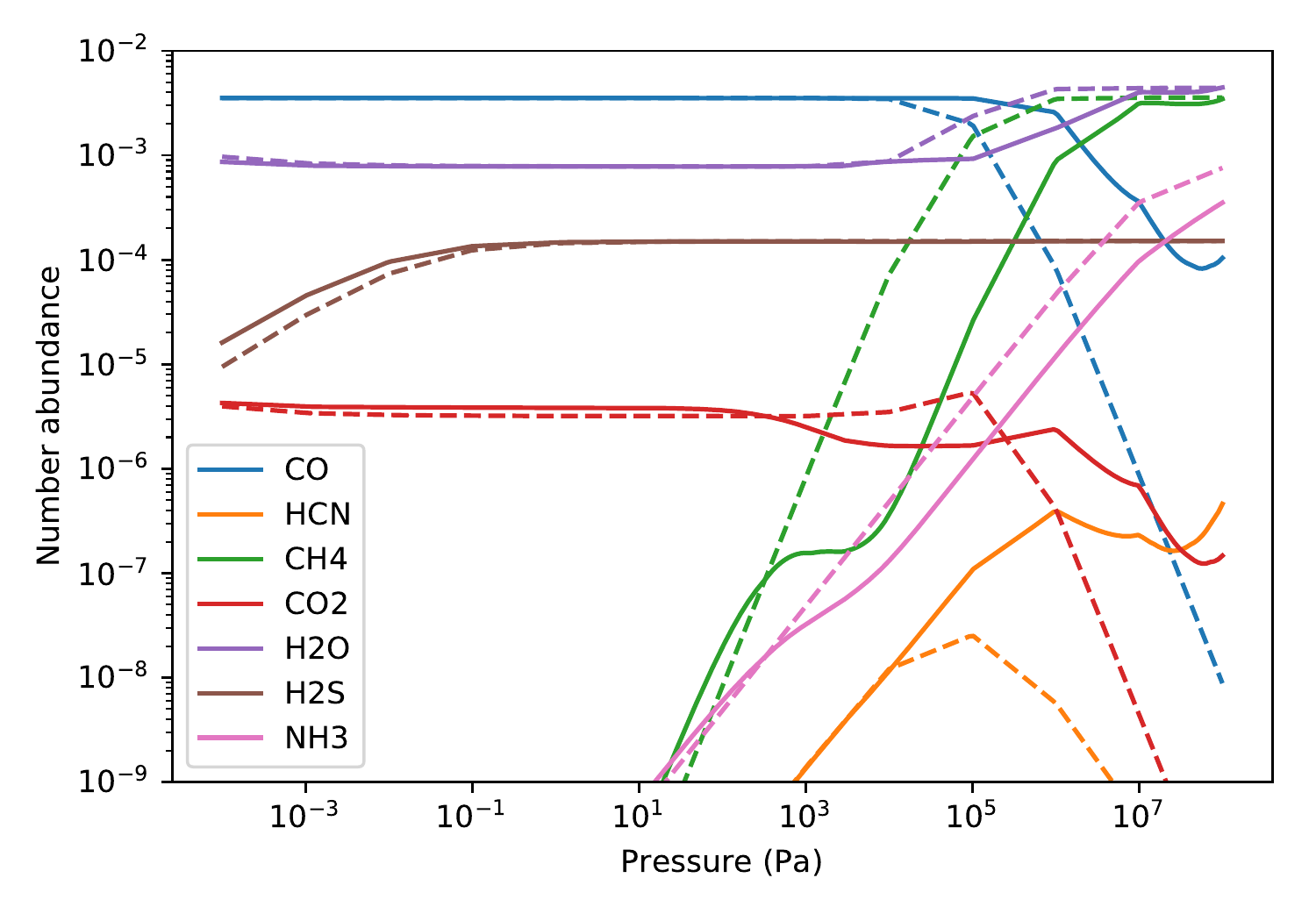}
    \caption{Number abundances of the most common molecules in the limb (dashed) and on the day side (solid), according to the best fit model.  Hydrogen and helium, the dominant components of the atmosphere, are not shown.}
\label{fig:abundances}
\end{figure}

In Table \ref{table:retrieved_parameters}, we tabulate the 1D posterior distributions from the retrieval.  In Figure \ref{fig:big_corner}, we show the 2D posterior distributions.  We find that the data prefer supersolar metallicities ($Z=7-20$) and a C/O ratio between $0.47-0.69$.  This C/O ratio is consistent with the solar value, but is somewhat low compared to the stellar C/O ratio of $0.90 \pm 0.15$ \citep{teske_2014}.  However, \cite{teske_2014} also report that their estimate for the stellar C/O ratio depends on what data they include and how the non-LTE correction is performed.  They report C/O ratios ranging from 0.69 to 1.2 for different data analysis choices, and conclude that although the C/O ratio could be below 0.75, it is very likely above 0.80.  If so, the planetary atmospheric C/O ratio would be slightly suppressed relative to the stellar value, in good agreement with theoretical predictions for gas giant planets with atmospheric metallicities enhanced by the accretion of solids \citep{espinoza_2017}.

Our observations also constrain HD 189733b's dayside pressure-temperature profile.  We find no evidence for a dayside temperature inversion, as shown in Figure \ref{fig:contrib_and_TP}.  The parameter $\alpha$, which partitions the visible radiation into two separate channels and corresponds to the flux in the channel with opacity $\gamma_2$, is consistent with 0. This means that there is no need for a second visible wavelength channel and that the simpler double-gray parameterization of \cite{guillot_2010} is sufficient. We also find that the overall shape of the pressure-temperature profile is consistent with a low albedo and efficient day-night redistribution of heat, as $\beta$ is consistent with 1.

We place constraints on the sizes and locations of the scattering particles near HD 189733b's terminator.  The mean particle size is constrained to be less than 14 nm, and is consistent with arbitrarily small values.  However, this size constraint is dependent on the assumed value of the imaginary refractive index, with more absorbent particles requiring a larger mean particle size (see Subsection \ref{subsec:aerosol} for more details).  We find that the fractional scale height of the haze is a factor of a few larger than that of the gas.  In effect, this means that haze particles are more abundant relative to gas in the upper atmosphere than in the lower atmosphere.  This could possibly indicate that the haze is photochemical in nature, a possibility first suggested by \cite{zahnle_2009} and \cite{pont_2013}.  The fact that photochemical hazes can generate super-Rayleigh scattering slopes in the $T_{eq}=1000-1500$ K regime was recently demonstrated by \cite{ohno_2020}, who showed that strong eddy diffusion gives rise to a $\rho_{\rm haze}/\rho_{\rm gas}$ ratio that increases with height, which in turn leads to a steep spectral slope.

\subsection{Comparison with previous retrievals}\label{sec:comparison}
Many authors have attempted to use retrievals to constrain the atmospheric properties of HD 189733b, starting with \cite{madhusudhan_2009}.  Here, we review the most recent retrievals, including \cite{lee_2014} and \cite{pinhas_2019} in transmission and \cite{lee_2012} in emission, and compare their results to ours. Although \cite{benneke_2015} separately performed a retrieval on HD 189733b's WFC3 transit spectrum, the limited wavelength range of these data prevented them from obtaining meaningful constraints on the atmospheric metallicity.

\cite{lee_2014} performed a retrieval on the $0.3-10$ \um transmission spectrum data reported in \cite{pont_2013}. Because this study was performed prior to the publication of the WFC3 data, it used NICMOS spectroscopy to constrain the shape of HD 189733b's near-infrared transmission spectrum.  As part of this retrieval they explored several different potential aerosol species, including MgSiO$_3$, Mg$_2$SiO$_4$, astronomical silicate (a mixture of siliciate grains commonly seen in interstellar space), MgSiO$_3$, NaS, and tholins, all of which they argued might plausibly form in the upper regions of HD 189733b's atmosphere.  They found that a vertically uniform aerosol layer stretching from 0.1 mbar to 10 bar filled with monodispersed particles smaller than 0.1 \um provides a good fit to the spectrum for all aerosol compositions.  This matches well with our updated fits, which indicate that the data are consistent with a large range of imaginary refractive indices and favor particles with a mean radius smaller than 14 nm, distributed with a scale height much larger than that of the gas.  \cite{lee_2014} also reported a constraint of $0.02-20 \times 10^{-4}$ on the abundance of water after accounting for the uncertainties introduced by different aerosol assumptions. Our best fit model is on the upper end of this range with a photospheric water abundance of $10^{-3}$, but it is consistent with the water abundances from this study for both Mg$_2$SiO$_4$ and tholin aerosols.

\cite{pinhas_2019} use AURA to perform a retrieval on the transit spectrum from STIS, ACS, WFC3, and IRAC 3.6/4.5 (0.35--4.5 \um).  They obtain a water abundance of $\log(X_{H_2O}) = -5.04_{-0.30}^{+0.46}$, which is 1.8\% the equilibrium value at solar elemental abundances, and claim a strong detection of water depletion.  We computed the water abundance at 10 mbar in our models, taking this to be representative of the photospheric pressure, and obtained $\log(X_{H_2O}) = -2.5 \pm 0.3$.  Our result is substantially discrepant with \cite{pinhas_2019}: it is marginally super-solar (see their Figure 2) and comparable to the measured atmospheric C/H ratio for Jupiter (e.g., \citealt{lodders_2003}, their Figure 6).  

The cause of this discrepancy is not clear.  In terms of data, \cite{pinhas_2019} include only the 3.6 and 4.5 \um Spitzer transit depths, while we also include transit depths at 5.8 \um, 8.0 \um, and 24 \um. The most significant difference in the methodology is that \cite{pinhas_2019} allows much more freedom than our retrieval.  We use equilibrium abundances, while they fit for the abundances of 6 individual atoms and molecules.  We adopt an isothermal limb, while they adopt the \cite{madhusudhan_2019} parameterization of the limb T/P profile and fit for all 6 parameters.  In total, they have 19 free parameters, while our transit-only retrieval has only 9.  The high number of free parameters in \cite{pinhas_2019} may allow them to find a better fit to the data, one with sub-solar water abundance.  On the other hand, the flexibility also puts them in more danger of over-fitting and of finding physically unrealistic compositions or T/P profiles.

We next consider previously published fits to HD 189733b's dayside emission spectrum.  \cite{lee_2012} used optimal estimation to perform a retrieval on all published eclipse observations, with wavelengths ranging from $1.45-24$ \um.  They did not include the WFC3 eclipse observations \citep{crouzet_2014}, which were published after that study, and instead included NICMOS observations spanning a similar wavelength range \citep{swain_2008}.  Using these data, they found a mixing ratio of $0.9-50 \times 10^{-4}$ for water, $3-150 \times 10^{-4}$ for carbon dioxide, and $< 0.4 \times 10^{-4}$ for methane, implying a C/O ratio of 0.45--1.  The error ranges they derived for CO were so broad that they could not provide meaningful estimates of its abundance.  Comparing to Figure \ref{fig:abundances}, we see that our water abundance of $10^{-3}$ is fully consistent with these results, as is our low methane abundance of $\sim10^{-7}$ at $P\sim0.1$ bar.  However, our model has several times less CO$_2$.  \cite{lee_2012} observed that previous studies preferred much smaller CO$_2$ abundances, including \cite{line_2010} ($10^{-7}$ to $10^{-5}$), \cite{swain_2009} ($10^{-7}$ to $10^{-6}$), and \cite{madhusudhan_2009} ($7-700 \times 10^{-7}$).  They concluded that it is their \emph{HST}/NICMOS data that caused the fits to prefer a high CO$_2$ abundance.  As discussed above, some studies have questioned the reliability of the NICMOS results \citep{gibson_2012, deming_seager_2017}, which also sometimes appear to contradict subsequent WFC3 observations \citep{deming_2013}.  A lower CO$_2$ abundance would also be more physically plausible, as equilibruim chemistry predicts that it should be relatively rare at the low atmospheric metallicities preferred by our model, and disequilibrium models including both photochemistry and quenching do not appreciably increase the predicted CO$_2$ abundance (e.g., \citealt{moses_2013, steinrueck_2019}.  

\begin{figure}[h]
  \centering 
  \includegraphics[width=\linewidth]{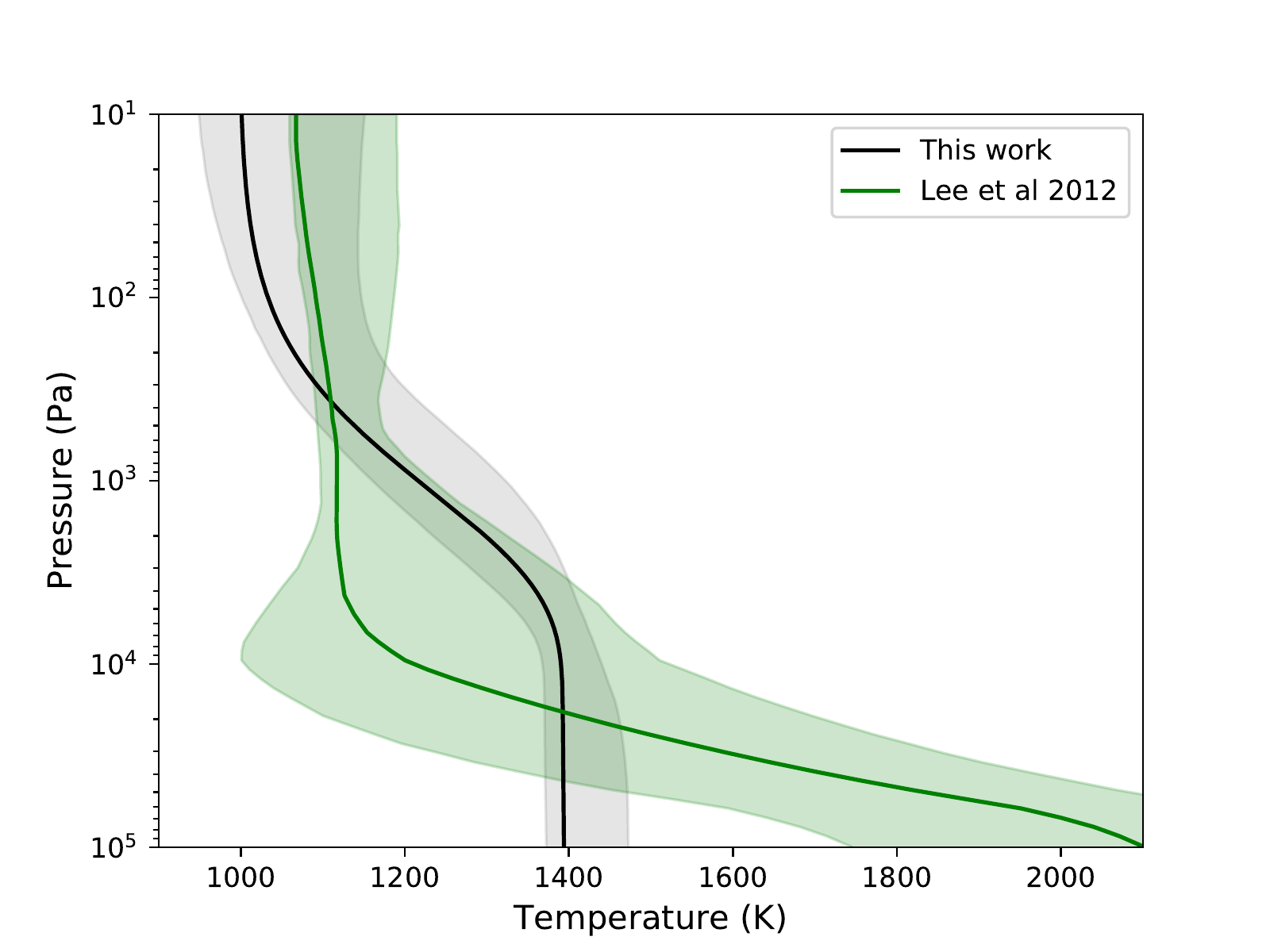}
  \caption{Comparison of our retrieved T/P profile (black) and that of \cite{lee_2012} (green).}
\label{fig:TP_comparison}
\end{figure}

In addition to abundances, \cite{lee_2012} also retrieved a T/P profile (their Figure 1), which we compare to our T/P profile in Figure \ref{fig:TP_comparison}.  Although the two profiles are very discrepant at higher pressures ($>\sim$1 bar), it is important to note that constraints on the T/P profile at pressures higher than 1 bar are imposed by the finite range of shapes allowed by the T/P profile parameterization, not by observational data.  This is because emission spectroscopy cannot probe those depths, as shown by the contribution function (Figure \ref{fig:contrib_and_TP}).  At low pressures the two profiles are consistent, despite the different shapes.

\subsection{Importance of individual molecules}
\platon calculates the abundances and line opacities of 28 molecules.  To ascertain which molecules are important, we re-calculate the best fit transit and eclipse spectra with the opacities of individual molecules zeroed out to see how ${\chi}^2$ changes.  We find that the transit spectrum is dominated by opacity from H$_2$O, CO$_2$, H$_2$S, and CH$_4$.  Removing all other molecular opacities only increases $\chi^2$ by 0.7.  Starting from a reference point consisting of these four molecules, we remove each molecule in turn and calculate the resulting $\Delta \chi^2$, obtaining 1.4 for H$_2$S, 2.9 for CO$_2$, 3.6 for CH$_4$, and 92.9 for H$_2$O.  Zeroing the aerosol opacity yields $\Delta \chi^2=3780$.  We conclude that water is by far the dominant molecule shaping the transit spectrum, with aerosol scattering as the most important opacity source overall.  This is no surprise, as both the strong scattering slope at short wavelengths and the water feature in the WFC3 bandpass are obvious by eye (Figure \ref{fig:best_fit_transits}).

The eclipse spectrum worsens by only $\Delta \chi^2=0.6$ when the opacity is zeroed for all molecules except H$_2$O, CO$_2$, H$_2$S, CH$_4$, and CO.  Starting from a reference point consisting of these five molecules, we remove each molecule in turn and calculate the resulting $\Delta \chi^2$, obtaining 0.2 for CH$_4$, 8.5 for H$_2$S, 56.6 for CO, 81.4 for CO$_2$, and 551 for H$_2$O.  Thus, we conclude that the emission spectrum contains information on more molecules than the transit spectrum, with H$_2$S, CO, CO$_2$, and H$_2$O all acting as important opacity sources, although water is still dominant.

The tests above reveal the contribution of different molecules to the best-fit transit and eclipse spectra.  They do not reveal the significance with which individual molecules are detected in the fits, because many of the features induced by a molecule--a little more absorption here, a little less absorption there--can be mimicked by changes in the free parameters.  To quantify the detection significance, we ran a series of retrievals on the transit and eclipse data where we zeroed out the opacity of one molecule at a time and calculated the resulting Bayesian evidence $z$.  The log of the Bayes ratio (indicating the relative preference for the full model versus one without that molecule) is then given by the difference in $\ln(z)$ when compared to the retrieval where all molecular opacities were included.  In transit, $\Delta \ln(z)$ was -0.9 for H$_2$S, 0.5 for CH$_4$, -0.8 for CO$_2$, and -9.9 for H$_2$O, with a margin of error on $\ln(z)$ equal to $\sim$0.2 for all retrievals. We conclude that the transit spectra only provide strong evidence for H$_2$O, with a Bayes factor of 20,000; all other molecules have a Bayes factor less than 3.  In eclipse, $\Delta \ln(z)$ was 0.3 for H$_2$S, -0.7 for CO, -1.2 for H$_2$O, and -1.6 for CO$_2$, with a similar margin of error.  We conclude that the eclipse spectrum does not strongly favor the existence of any one molecule.  These results are consistent with intuition: there is a visually obvious water feature in the WFC3 transit spectrum, but no molecular features can be seen in the eclipse spectrum (which is predominately composed of broadband photometric points) at any wavelength.

\subsection{Aerosol properties}
\label{subsec:aerosol}
Many authors have proposed a Rayleigh scattering haze to explain the optical transmission spectrum of HD 189733b (e.g. \citealt{lecavelier_2008, gibson_2012, pont_2013}), and this is born out by our retrievals.  Our posteriors indicate that a clear atmosphere with a zero number density of haze particles is ruled out to much greater than $3\sigma$ significance, and the posterior distribution of particle sizes puts the particles firmly in the Rayleigh regime ($r \lesssim \lambda/10$) for optical wavelengths.  In fact, our posterior on the mean particle sizes pushes up against 1 nm, the lower end of the prior--indicating that arbitrarily small particles are allowed by the data.  This means that no constraint on the haze composition is possible, as the scattering slope is always $\frac{dR_p}{d\log{\lambda}} = -4H$ regardless of composition.  The lack of a lower limit on the mean particle size also implies that there is no upper limit on the particle density, as there is a perfect degeneracy between the two variables in the Rayleigh regime.

Our findings are consistent with the conclusions of previous studies (i.e. \citealt{gibson_2012,pont_2013}), which required the inclusion of a Rayleigh scattering haze in order to reproduce HD 189733b's infrared transmission spectrum.

\begin{figure}[ht]
  \centering 
  \includegraphics[width=\linewidth]{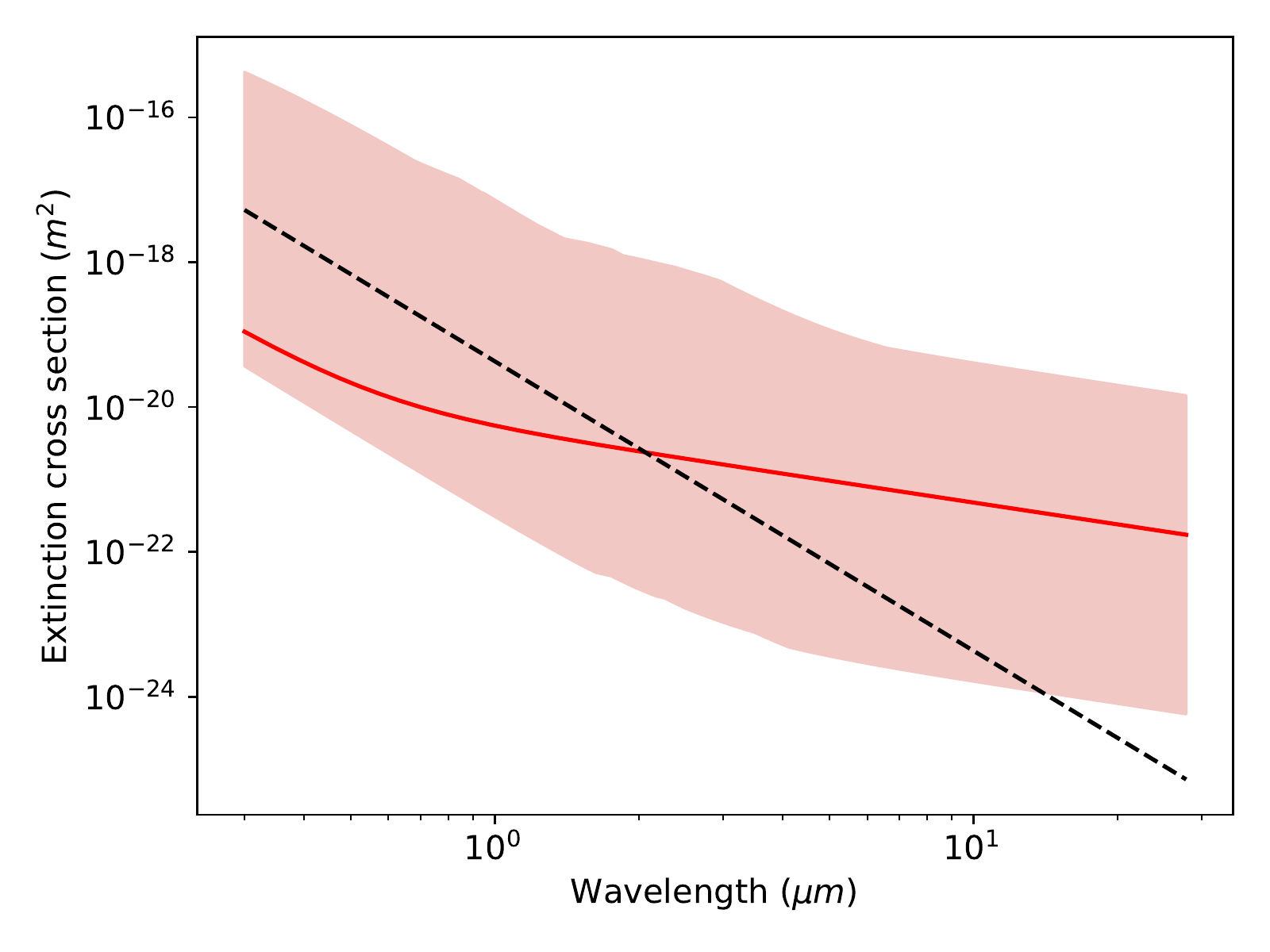}
  \caption{Extinction cross section of the haze particles in our best fit model (solid red), along with the 1$\sigma$ range of extinction cross sections from our retrieval.  The dashed solid line falls off as $\lambda^{-4}$, and is plotted for reference.}
\label{fig:mie_extinction}
\end{figure}

In Figure \ref{fig:mie_extinction}, we show that the aerosol extinction cross section falls as $\lambda^{-4}$ at very short wavelengths, but shifts to $\lambda^{-1}$ at longer wavelengths.  This slow dropoff of extinction cross section at high wavelengths is due to aerosol absorption.  To understand this, we can write down simple analytic expressions for scattering and absorption cross sections as a function of wavelength in the Rayleigh regime.  These are \citep{mishchenko_2002}:

\begin{align}
    \sigma_{sca} &= \frac{2^7 \pi^5 r^6}{3\lambda^4} \abs*{\frac{m^2 - 1}{m^2 + 2}}^2\\
    \sigma_{abs} &= \frac{8 \pi^2 r^3}{\lambda} Im\Big(\frac{m^2 - 1}{m^2 + 2}\Big)
\end{align}
Writing the complex refractive index $m$ as $n$ + i$k$, and assuming $k << n$:

\begin{align*}
    \sigma_{sca} &= \frac{2^7 \pi^5 r^6}{3\lambda^4}\frac{4k^2n^2 + (n^2-k^2-1)^2}{4k^2n^2 + (n^2-k^2+2)^2}\\
    &\approx \frac{2^7 \pi^5 r^6}{3\lambda^4}\frac{(n^2 - 1)^2}{(n^2 + 2)^2}\\
    \sigma_{abs} &= \frac{8 \pi^2 r^3}{\lambda}\frac{6nk}{(n^2-k^2+2)^2 + 4k^2n^2}\\
    & \approx \frac{8 \pi^2 r^3}{\lambda}\frac{6nk}{(n^2+2)^2}
\end{align*}

One can see from these equations that aerosol scattering falls off with wavelength much faster than aerosol absorption, causing extinction to be dominated by scattering at short wavelengths and absorption at long wavelengths.  Since the absorption cross section is proportional to the imaginary component of the refractive index, even a small imaginary component increases long wavelength extinction by orders of magnitude compared to a real refractive index.

The importance of $k$, the imaginary component of the refractive index, poses a challenge for our model.  Many different cloud species have been proposed for this planet.  \cite{lee_2016} used a GCM simulation to model condensate clouds and found that the clouds are dominated by silicate materials such as MgSiO$_3$ at mid-high latitudes, but TiO$_2$ and SiO$_2$ dominate in equatorial regions.  \cite{lavaas_2017} considered photochemical hazes and found that soot-composition aerosols provided a good match to HD 189733b's transmission spectrum.  This study reported $k\sim$ 0.5 at 500 nm for soot, while \cite{kitzmann_2018} reported 3.7\e{-5} for glassy MgSiO$_3$ at the same wavelength, 5.1\e{-4} for TiO$_2$, and 1.7\e{-5} for SiO$_2$.  In addition to these species-dependent variations in $k$, the $k$ for each species also varies drastically (and uniquely) with wavelength.  For example, $k$ rises from 1\e{-4} to nearly 1 over the wavelength range $2-9$ \um for MgSiO$_3$, and SiO$_2$ exhibits a similar behavior before dropping 2 orders of magnitude within 2 \um.  In light of these uncertainties, we chose to fit for a wavelength-independent $k$ value rather than fixing it to the theoretical prediction for a given cloud species.

We argued earlier that the presence of aerosol has a negligible effect on HD 189733b's dayside emission spectrum.  We check the validity of this assumption using our best-fit model.  We find that when we include the best-fit aerosol model from the transmission spectrum in our calculation of HD 189733b's dayside emission spectrum the resulting eclipse depth values change less than $\sim$0.1 ppb, with a corresponding change in ${\chi}^2$ of only 2\e{-5}.  This is unsurprising, as the photospheric pressure is lower for transmission spectrum than it is for eclipse spectrum by a factor of $\sqrt{2\pi R/H} \sim 50$ \citep{fortney_2005}; this is apparent when we compare the transmission spectrum contribution function in Figure \ref{fig:transit_contrib} to the emission spectrum contribution function in Figure \ref{fig:contrib_and_TP}.  At higher pressures, the mixing ratio of aerosol is lower because our retrievals prefer an aerosol scale height that is greater than the gas scale height.  This causes aerosol absorption to be an important source of opacity at low pressures only.  We note that the dayside is also expected to be hotter than the terminator, making it less likely that the condensate clouds detected at the terminator would persist in this region.  Even if the dayside is in reality partly cloudy, the dayside emission would be dominated by clear regions because they have deeper and hotter photospheres, which emit more radiation.



\subsection{Validity of equilibrium chemistry}
\platon assumes equilibrium chemistry.  Since chemical reaction timescales decline rapidly with temperature, the colder a planet, the more disequilibrium chemistry matters.  HD 189733b lies in a regime where disequilibrium chemistry may be important.  Multiple studies have explored disequilibrium chemistry on this planet \citep{line_2010,venot_2012,moses_2013,agundez_2014,blumenthal_2018,steinrueck_2019}.   \cite{venot_2012} considered 1D models with both UV photochemistry and vertical mixing, and found negligible differences in the resulting transmission and emission spectra as compared to equilibrium models.  Their Figure 10 shows that the disequilibrium-induced brightness temperature discrepancy is at most several Kelvin (less than 1\%), while the transit radius discrepancy is at most $\sim$30 ppm.  \cite{blumenthal_2018} also model the emission spectrum under equilibrium and non-equilibrium conditions, finding no detectable difference even with JWST (their Figure 3). Other studies report changes in abundance as a result of disequilibrium chemistry, but do not compare the resulting spectra to the equilibrium model predictions.  For example, \cite{moses_2013} find that the water abundance only becomes discrepant for P $<$ 1 microbar, while the HCN abundance is enhanced by orders of magnitude for P $<$0.5 bar.  In contrast to these 1D models, which predict relatively small changes in HD 189733b's observed transmission and emission spectrum, \cite{steinrueck_2019} calculated emission spectra for HD 189733b using general circulation models where they fixed the ratio of CH\textsubscript{4} to CO across the planet to mimic the effect of transport-induced quenching.  They find that disequilibrium chemistry due to horizontal transport changes the emission spectrum by up to 20\% for a heavily CO-dominated atmosphere (CH\textsubscript{4}/CO = 0.001): there is a systematic offset of 10\% in addition to wavelength-dependent discrepancies of order several percent.

To test whether disequilibrium chemistry may be important for our data, we performed retrievals on the transit and eclipse spectra with the vertical quench pressure as a free parameter.  We quenched all molecules in our first trial, but only CH4 and CO in our second trial (in accordance with \citealt{morley_2017}), with no change in the following conclusions.  Quenching did not result in a better fit ($\Delta \chi^2 \sim 0$) for either our transmission or emission spectra.  In both cases, the posterior distribution of the quench pressure pushes up against the lower bound of the prior--0.1 mbar for the transit spectrum and 1 mbar for the eclipse spectrum.  Additionally, the posterior distributions of the other parameters did not appreciably change in these retrievals.  For example, the metallicity posterior shifted by 0.1$\sigma$ for the transit retrieval and 0.5$\sigma$ for the eclipse retrieval; the C/O ratio posterior shifted by 0.06$\sigma$ for both transit and eclipse retrievals.  We conclude that the effects of disequilibrium chemistry are below our detection threshold, in good agreement with the predictions of the 1D models from \cite{venot_2012} and \cite{blumenthal_2018}.

\subsection{The effect of starspots}
\label{subsec:activity}
HD 189733 is an active K dwarf with spots that cover a few percent of its surface.  When the planet crosses a prominent spot it creates a readily identifiable deviation in the transit light curve shape; this makes it straightforward to identify and mask such events in high signal-to-noise \emph{HST} observations \citep[e.g.][]{sing_2011,pont_2013}.  However, the presence of unocculted spots during a transit also biases the retrieved transit depth in a way that can mimic the effect of scattering as discussed in \cite{mccullough_2014}. If the fractional spot coverage varies from one transit epoch to another, the relative effect on the transit depth will vary as well.  An additional complication is faculae, bright regions $\sim$100 K hotter than their surroundings. For a K1.5 star like HD 189733, faculae can cover 17-40\% of the surface, partially cancelling the effect due to spots \citep{rackham_2019}.  The large faculae fraction is a mixed blessing.  On one hand, it becomes hard to correct for the effects of faculae because we cannot assume faculae are unocculted. On the other hand, the faculae covering fraction is so large that the planet is likely to cross at least one faculae and one non-faculae region during every transit, thus mitigating the spectral bias from unocculted faculae.  We therefore neglect the effects of faculae and only consider spots.

\cite{pont_2013} attempted to correct for the effects of time-varying spot coverage by using photometric data from the Automated Patrol Telescopes \citep{henry_1999} to estimate the apparent brightness at the epoch of each transit observation.  Although these data made it possible to estimate the magnitude of the variations in spot coverage, they do not provide an estimate of the overall spot coverage fraction because the star is not necessarily spotless at maximum brightness.  The authors resolve this problem by assuming an average spot coverage fraction of 1\%, based on three pieces of evidence: the frequency of spot-crossing events during HST transits \citep{sing_2011}, stochastic starspot simulations by \cite{aigrain_2012}, and the lack of features in the transmission spectrum (i.e. Mg H line, stellar sodium line) caused by abundant starspots.  They argue that these lines of evidence make it unlikely that the spot coverage fraction is much above 2\%.

\cite{mccullough_2014} use the same data to argue for a higher starspot fraction of 4\%.  First, they emphasize that the starspot fraction derived from spot crossing events would be an underestimate if most starspots are in the polar regions, where the planet does not transit.  Second, they use Equation 14 of \cite{aigrain_2012} to derive a lower bound on the starspot fraction:

\begin{align*}
    \delta >\approx \frac{\Psi_{max} - \Psi_{min} + \sigma}{\Psi_{max} + \sigma}
\end{align*}
Since the measured difference between the maximum flux ($\Psi_{max}$) and minimum flux ($\Psi_{min}$) is around 4\% and the rotational modulation ($\sigma$) is of order one percent, $\delta >\approx 0.04$.  However, this derivation really computes a lower bound on the maximum starspot fraction over a period of time, whereas the 1-2\% figure quoted by \cite{pont_2013} is the average starspot fraction. \cite{mccullough_2014} go on to argue that if the starspot fraction were 4.3\%, the majority of the increased apparent transit depth in the UV compared to IR could be explained by unocculted starspots, and there would be no need to invoke scattering from aerosol.

To shed light on this issue, we ran another retrieval with \platon where the spot coverage fraction is allowed to vary as a free parameter.  We fix the spot temperature to 4250 K, the temperature derived by \cite{sing_2011} from observations of spot occultations.  Since we used the transit depths from \cite{pont_2013}, which already corrected for the effects of starspots using APT photometry and an assumed 1\% baseline starspot fraction, our fitted spot coverage fraction is in reality an excess spot fraction above this 1\% baseline. We find an excess spot coverage fraction of $1.8_{-1}^{+0.7}$\%, which in turn implies a total average starspot fraction of $2.8_{-1}^{+0.7}$\%.  This figure is intermediate between the 1-2\% argued for by \cite{pont_2013} and the 4\% argued by \cite{mccullough_2014}. However, the inclusion of spot coverage fraction as a free parameter does little to modify the posteriors for the other model parameters.  The median values of all parameters are consistent with the values from the fiducial retrieval (Table \ref{table:retrieved_parameters}) to better than 1$\sigma$.  We therefore conclude that our derived haze properties, including the mean particle size, particle number density, and fractional scale height, are insensitive to our assumptions about the spot coverage fraction.  

\begin{figure}[ht]
  \centering 
  \includegraphics[width=\linewidth]{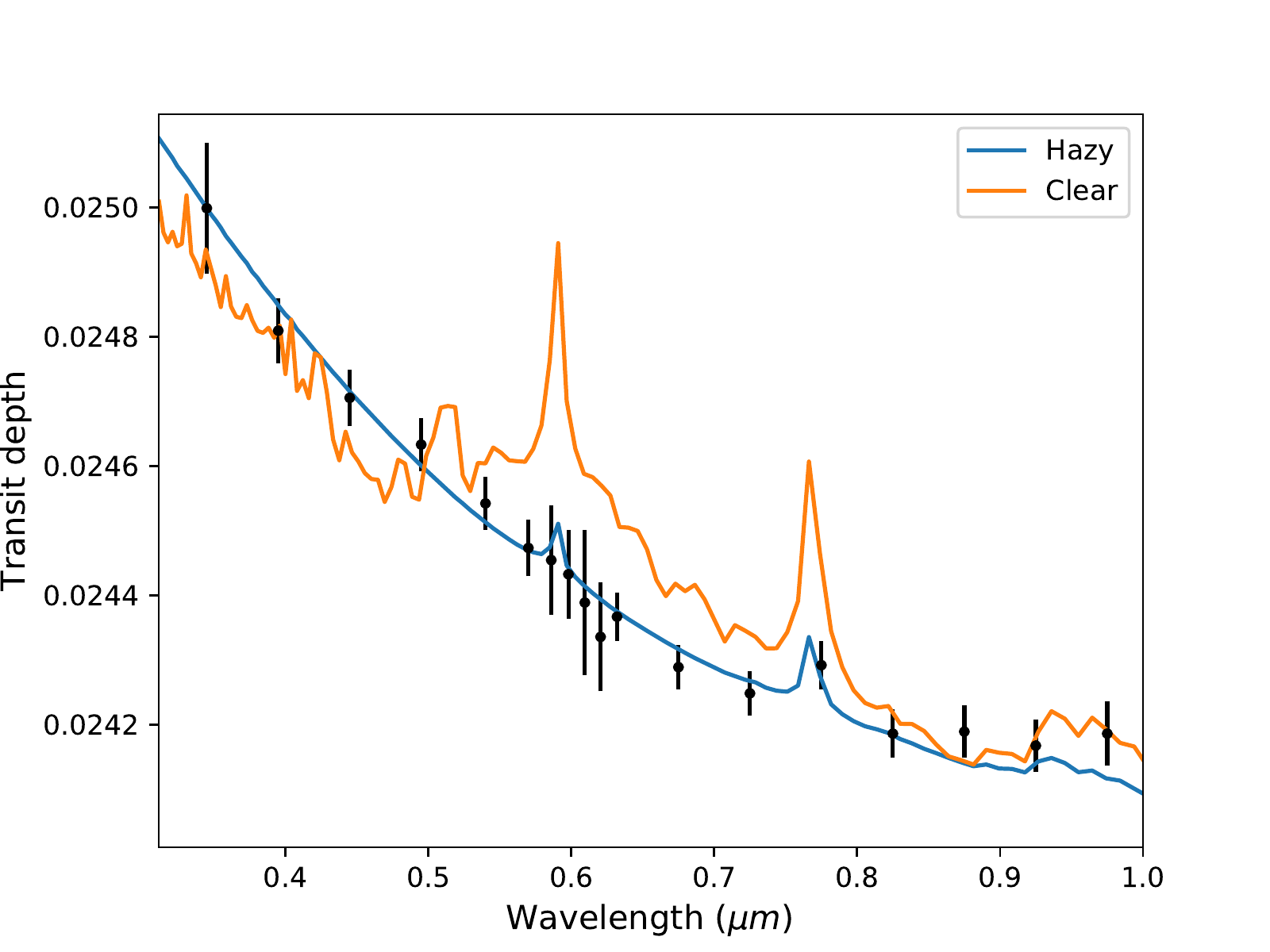}
  \caption{\platon model of a hazy atmosphere on HD 189733b, compared to a clear atmosphere with a starspot-induced slope corresponding to a spot coverage fraction of 6\%.  Note the strong atomic and molecular features in the clear atmosphere, especially the far wings of the sodium and potassium absorption lines.}
\label{fig:clear_atm}
\end{figure}

If unocculted starspots and haze can both introduce a slope in HD 189733b's optical transmission spectrum, why is there not a degeneracy between the two?  To answer this question, we plotted a transmission spectrum with no aerosol but with a high spot coverage fraction of 6\%, as shown in Figure \ref{fig:clear_atm}.  We found that the transmission spectrum has strong atomic and molecular absorption features even at short wavelengths ($\lambda < 1$ \um), the most prominent of which are the far wings of the Na doublet at 589 nm and of the K doublet at 770 nm.  These features are not seen in the observational data, which are nearly featureless at optical wavelengths.

\subsection{A consistency check for composition}
\begin{figure}[ht]
  \centering 
  \includegraphics[width=\linewidth]{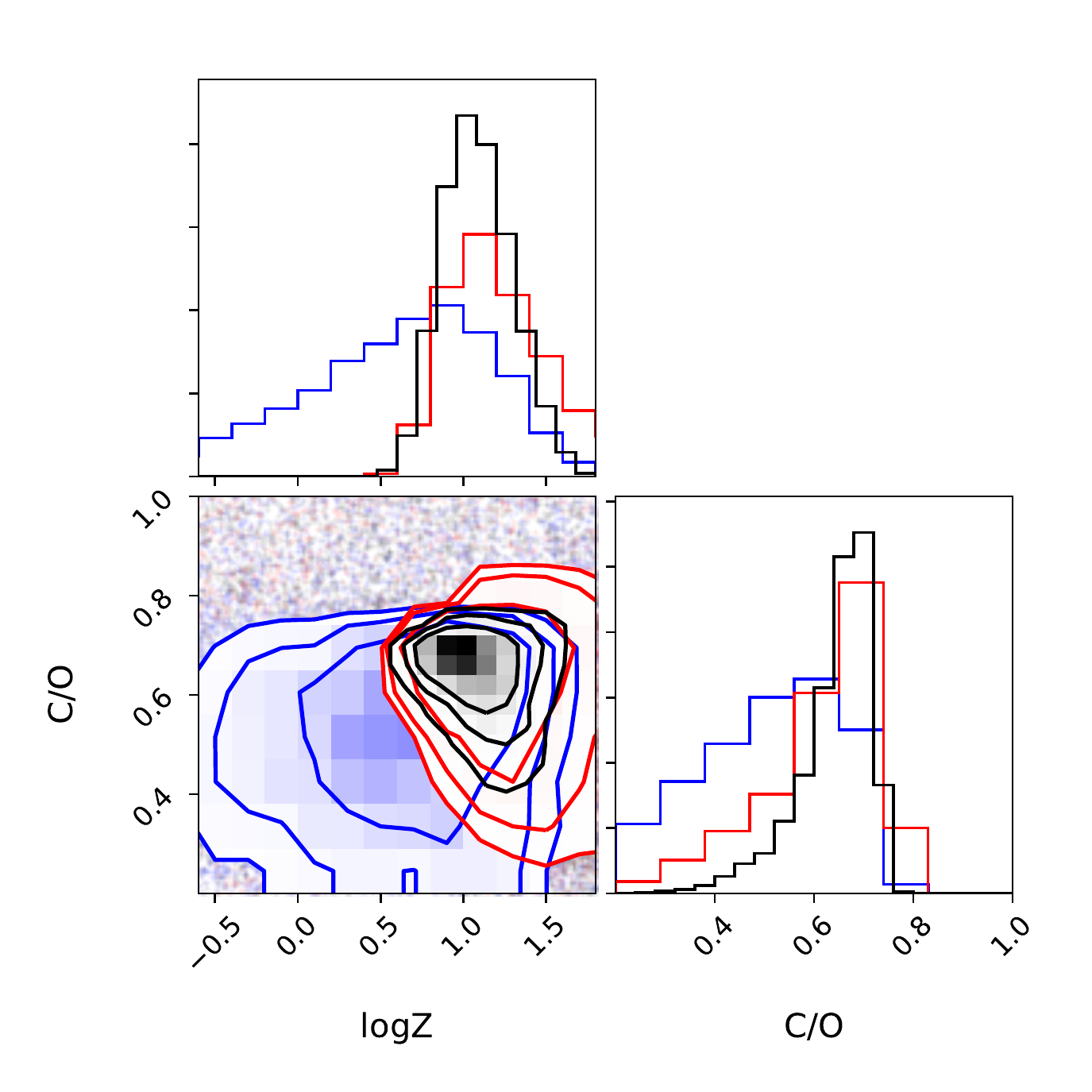}
  \caption{Posterior distributions for the transit-only retrieval (blue), eclipse-only retrieval (red), and combined retrieval (black).}
\label{fig:transit_vs_eclipse_corner}
\end{figure}

Although we fit HD 189733b's transmission and emission spectra jointly, in reality the models used to fit these two spectra are largely independent of each other.  The transmission spectrum determines the isothermal limb temperature and aerosol properties, while the emission spectrum sets the dayside T/P profile.  The planetary radius is technically constrained by both transmission and emission spectra, but most of the statistical power comes from the transmission spectrum.  The only two parameters that are comparably constrained by both spectra are the atmospheric metallicity and C/O ratio, which determine the chemistry for both the limb and the day side.

To illustrate the relative contributions of transmission and emission spectra to HD 189733b's inferred atmospheric composition, we ran a transit-only and an eclipse-only retrieval.  In the transit-only retrieval, the planetary radius, metallicity, C/O ratio, isothermal limb temperature, aerosol properties, and WFC3 offset were free parameters.  In the eclipse-only retrieval, the metallicity, C/O ratio, dayside T/P profile parameters, and WFC3 offset were free parameters.  Figure \ref{fig:transit_vs_eclipse_corner} shows the resulting posterior distributions for metallicity and C/O from these two retrievals, with the distributions from the combined retrieval overplotted.  The three retrievals give fully consistent constraints on both parameters.  The emission spectrum puts a slightly tighter constraint on metallicity, but both spectra place comparable constraints on the C/O ratio.

\subsection{High resolution studies}
\label{subsec:hd189_high_res}
HD 189733b has been a favorite target for high resolution spectroscopy.  VLT/CRIRES detected water on its dayside \citep{birkby_2013}, as well as both water and CO on its terminator \citep{brogi_2016}. TNG/GIANO and CARMENES also detected water in the transit spectrum \citep{brogi_2018,alonso-floriano_2019}.  Carbon monoxide has been detected on its dayside by both Keck/NIRSPEC \citep{rodler_2013} and VLT/CRIRES \citep{de_kok_2013}.  The dayside detections with CRIRES were confirmed by \cite{cabot_2019}, who additionally report a high confidence (5.0$\sigma$) detection of HCN, with a HCN mixing ratio of $10^{-6}$ yielding peak detection significance.

\begin{figure}[h]
  \centering 
  \includegraphics[width=\linewidth]{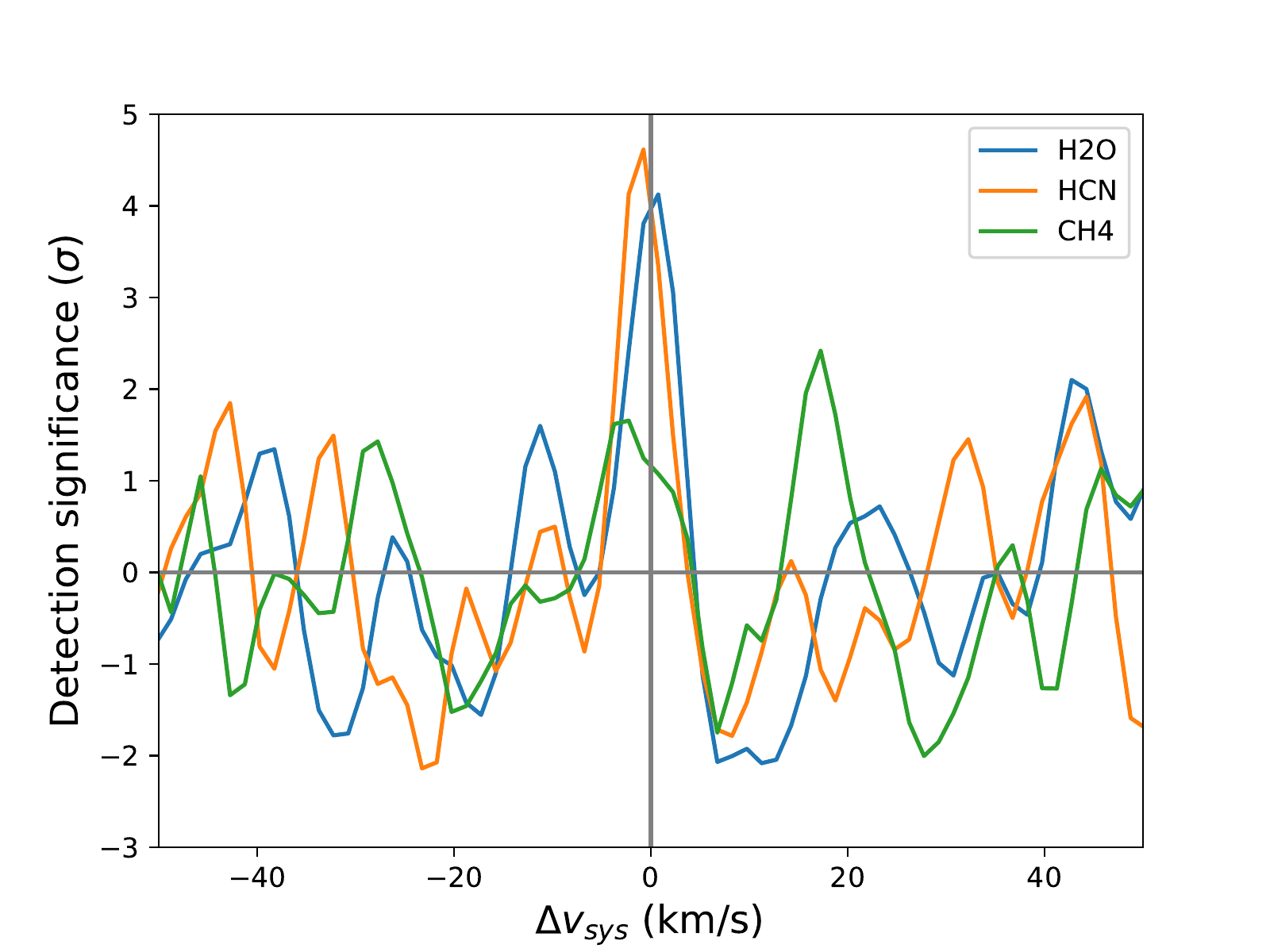}
  \caption{Detection significance of H$_2$O, HCN, and CH$_4$ in the high resolution emission spectrum of HD 189733b.  H$_2$O and HCN are seemingly clearly detected, while CH$_4$ is not.  The model used to cross correlate with data was generated by \platon using the parameters of the best fit model from the low resolution retrieval.  Despite the seeming robustness of the HCN detection, bootstrap analysis reveals that its actual significance is 1.5--1.9$\sigma$.}
\label{fig:detection_significances}
\end{figure}

We test the robustness of the water and HCN detections with \platon.  We replicate the methodology in \cite{cabot_2019} to reduce the CRIRES L band (3.18--3.27 \um) data.  We then generate a high-resolution \platon model of HD 189733b assuming the best fit parameters to the low resolution data, with only the line opacity of the molecule in question included.  The model eclipse spectrum is cross correlated with the data to look for a signal with the expected radial velocity drift of the planet.  We detect H$_2$O and HCN at a significance of 4.3$\sigma$ and 4.8$\sigma$ respectively, in line with the results of \cite{cabot_2019}.  These results are shown in Figure \ref{fig:detection_significances}.

While both molecular detections seem robust, they may not be.  Standard high resolution analysis methods involve optimizing many parameters to maximize the detection significance.  For \cite{cabot_2019}, these are the number of SYSREM (principal component analysis to remove telluric features) iterations for each of the 4 detectors, the planetary orbital velocity, the systemic velocity offset, the percentage of wavelengths to mask due to low atmospheric transmission, and the percentage of wavelengths to mask due to high variability.  They report from injection-recovery tests that this optimization procedure yields false positives as high as 4.0$\sigma$ 30\% of the time.  

To reduce the bias introduced by the optimization procedure, we fix the orbital velocity to 152.5 km/s \citep{brogi_2016} and restrict the systemic velocity offset to $\pm 1$ km/s.  We then estimate the bias introduced by the optimization process by a bootstrap-inspired procedure.  We randomly select, with replacement, 48 spectra from the original list of 48 spectra to form a new list of 48 spectra in random order.  We then apply the same analysis used for the original set of spectra to the new set.  This random selection and ordering, combined with the radial acceleration of the planet (amounting to 2 pixels/spectrum), means that the lines in the template will rarely match up with the planetary absorption lines.  We expect that the detection significance of all molecules in this scenario should be $0\sigma$, and the magnitude of the recovered signals therefore allows us to estimate the bias introduced by the optimization steps. 

We run this bootstrap procedure 1000 times, gathering the ``detection significance'' from each run.  We find that our optimization procedure, despite exploring fewer free parameters than \cite{cabot_2019}, returns an average bias of 1.1$\sigma$ for water and 2.9$\sigma$ for HCN.  Returning to the original analysis, our bias-corrected detection significance is then 3.2$\sigma$ for water, and 1.9$\sigma$ for HCN.  We conclude that the water detection is statistically significant, but the HCN detection is not.

In order to reduce the magnitude of this bias, we modified our analysis to minimize the size of the parameter space we optimize over.  Instead of optimizing the two masking parameters, we fix them to reasonable values: we mask wavelengths where the atmospheric transmission is under 30\%, and where the standard deviation of the wavelength across all spectra is in the top 10\% of all standard deviations for that detector.  Instead of optimizing over the four-dimensional space consisting of the number of SYSREM iterations for each detector, we optimize the number of iterations for each detector in turn before summing their weighted CCFs.  In addition, we include data from all detectors, instead of excluding the second detector due to the strong telluric absorption at those wavelengths.  

These changes turn a 6 dimensional optimization problem into four 1-dimensional optimization problems, each of which can have one of 10 discrete values, drastically decreasing the potential for bias.  We also apply a fourth order Butter high-pass filter to both the template and the data with a cutoff frequency of 0.01 pixel$^{-1}$, further reducing the potential for low-frequency systematics to create false signals.  These simplifications have the additional benefit of making our optimization code much faster.  In this new analysis, we find that water is detected at 3.8$\sigma$ with an optimization bias of 0.7$\sigma$ from bootstrapping, resulting in a corrected significance of 3.1$\sigma$.  HCN is detected at 2.8$\sigma$ with an optimization bias of 1.3$\sigma$, with a corrected significance of 1.5$\sigma$.  These results are in good agreement with our previous conclusions: the water detection is secure, but the HCN detection is statistically insignificant.

Why does our HCN measurement consistently exhibit higher levels of bias than water?  Although a detailed investigation of this question is outside the scope of this paper, we offer some speculations based on the differing statistical properties of the water and HCN line lists.  Water has a large number of weak lines irregularly spaced across the band, while HCN has a smaller number of very strong lines with a regular periodic spacing.  Thus, for HCN, the cross correlation function is dominated by strong lines spanning a smaller number of pixels, making overfitting more likely.  The cross correlation function for water depends on contributions from many pixels, making it more difficult to overfit.  

The periodic nature of the HCN lines means that the same problem occurs in Fourier space as in wavelength space.  Cross correlation is mathematically equivalent to taking the Fourier transform of the template, multiplying it by the conjugate of the Fourier transform of the data, and inverse Fourier transforming the product.  For templates with more periodic features, the Fourier transform of the template is dominated by a few high peaks.  These peaks may by chance coincide with unsubtracted periodic systematics, resulting in a spurious signal.  The Fourier transform for the water template is more evenly distributed, and is therefore less prone to this problem.

There are also physical reasons to doubt the HCN detection.  The detections of H$_2$O and CO are fully consistent with our best fit model to the low resolution data, which predicts that these molecules should be the most abundant active gases in the atmosphere at pressures lower than 1 bar.  These same models predict that the spectral signature of HCN should be significantly weaker than the signals from the more abundant H$_2$O and CH$_4$.  CH$_4$ has a slightly higher absorption cross section between $3.18-3.27$ \um, the wavelength range CRIRES covers.  While H$_2$O has a lower cross section over these wavelengths, it is many orders of magnitude more abundant.  Therefore, we would not expect HCN to be observable unless its abundance exceeds that of methane and is at least $\sim$10\% that of water; this would be many orders of magnitudes higher than predictions from equilibrium chemistry models for this planet.

One way around this difficulty is to invoke disequilibrium chemistry, as HCN abundances are enhanced both by transport-induced quenching and by photochemistry.  \cite{moses_2011} simulated HD 189733b and found that its HCN abundance is enhanced by disequilibrium processes.  Encouragingly, they found HCN abundances close to $10^{-5}$, similar to the $10^{-6}$ inferred observationally by \cite{cabot_2019}.  However, they find that CH$_4$ abundances are enhanced by the same processes, leaving the CH$_4$ abundance greater than or similar to the HCN abundance at typical photospheric pressures of 100 mbar.  They also find that the water abundance is not changed except at very high altitudes. 

The dominance of CH$_4$ in the models raises the question of whether the high-resolution CRIRES data shows any evidence of CH$_4$.  We searched for methane using the same CRIRES data by utilizing the methane line list by \cite{rey_2017}, which is complete at these high temperatures and has accurate line positions suitable for high resolution studies.  We find only a 1.2$\sigma$ signal, which is not significant.  This result is shown in Figure \ref{fig:detection_significances}.

\section{Conclusion}
\label{sec:conclusion}

A new and improved \platon is available for download.\footnote{Latest version: \url{https://github.com/ideasrule/platon}; version corresponding to this paper: \url{https://doi.org/10.5281/zenodo.3923090}}.  It now comes with an eclipse depth calculator, updated opacities, joint transit-eclipse retrieval capability, and correlated-$k$ capability.  In addition, we provide high resolution opacity data, making line-by-line calculations possible for the first time. We demonstrate \platon's new capabilities by using it to simultaneously analyze the best available \emph{HST} and \emph{Spitzer} transit and eclipse depths for the archetypal hot Jupiter HD 189733b.  To our knowledge, this is the first published retrieval on this comprehensive data set, as well as the first published joint retrieval that includes both transmission and emission spectroscopy.

Our resulting inferences for the properties of HD 189733b's atmosphere are qualitatively similar to--but more constraining than--those of previous authors.  We find that the data favors a haze with a mean particle radius less than 14 nm.  Our fiducial T/P profile indicates that the planet is consistent with a zero-albedo object with perfect heat redistribution.  We find that the atmosphere is of moderately super-solar ($7-21\times$ solar) metallicity and constrain the C/O ratio to lie between $0.47-0.69$, consistent with the solar value, but possibly lower than the stellar value.  This planet has one of the tightest metallicity constraints ever measured, with only WASP-127b and WASP-39b being comparable (see Figure 22 of \citealt{spake_2019}).  In our best-fit model, CO and H$_2$O are the most abundant absorbing species at photospheric pressures, consistent with the detection of H$_2$O in \emph{HST}/WFC3 spectroscopy, and the detection of both molecules in high resolution spectroscopy.

We explore the effects of stellar activity using a retrieval in which the starspot coverage fraction is allowed to vary as a free parameter, and find a best-fit starspot coverage fraction of $1.8_{-1}^{+0.7}$\%.  Even when this coverage is allowed to vary, our fit still requires the presence of a haze with much the same properties as the fiducial retrieval in order to create a featureless optical transit spectrum.

HD 189733b has exceptional observational data unmatched by any other exoplanet in quality, quantity, or wavelength range.  Much of this data was collected by Spitzer or by now-defunct instruments on HST, and can never be replicated.  Our retrieval demonstrates what kinds of properties can be inferred, and to what precision, for the most observationally favorable hot Jupiters in the pre-JWST era.  We have come close to testing the prediction by \cite{espinoza_2017} that enhanced atmospheric metallicity is inevitably associated with sub-stellar C/O ratio--in fact, the main obstacle was the uncertain stellar C/O ratio.  In addition, although the data for HD 189733b are already far more accurate than our models--the 3.6, 4.5, and 8.0 eclipse depths have errors of 1-2\%, whereas we have demonstrated that small changes in \platon's (and TauREx's) algorithm change the eclipse depth by several percent--we cannot fully take advantage of this accuracy in a retrieval.  This is because a retrieval has enough free parameters to fit the 3.6, 4.5, and 8.0 eclipse depths to arbitrary accuracy, and the lack of any molecular features in emission contributes to wide posteriors on atmospheric parameters despite the exceptional data.  JWST will be able to accurately measure multiple emission bands from multiple molecules across a large wavelength range, spurring spur the development of more sophisticated and more accurate models than the current state of the art.

\section{Acknowledgments}
M.Z. acknowledges Plato (Greek:$\Pi \Lambda\mathrm{AT}\Omega \mathrm{N}$) for his clear and thought-provoking dialogues, which ought to be exemplars of good writing for academics everywhere.  We also thank Michael R. Line for helpful advice.  Support for this work was provided by HST GO programs 13431, 13665, and 14260.

\textit{Software:} \texttt{numpy \citep{van_der_walt_2011}, scipy \citep{virtanen_2020}, matplotlib \citep{hunter_2007}, emcee \citep{foreman-mackey_2013}, dynesty \citep{speagle_2019}, corner \citep{foreman-mackey_2016}, nose, Travis-CI}

\section{Appendix}

The following HD 189733b parameters in Table \ref{table:real_planet_params} were fixed during the retrieval:

\begin{table}[ht]
  \centering
  \caption{Fixed planetary parameters in HD 189733b retrieval}
  \begin{tabular}{c c}
  \hline
  	Parameter & Value \\
      \hline
      $T_s$ & 5052 K\\
      $R_s$ & 0.751 $R_{\odot}$ \\
      $M_p$ & 1.129 $M_J$ \\
      a & 0.03142 AU\\
      \hline
  \end{tabular}
  \tablecomments{All parameters are taken from \citealt{strassun_2017} except the last, which is from \citealt{southworth_2010}}
  \label{table:real_planet_params}
\end{table}

Tables \ref{table:adopted_transit_depths} and \ref{table:adopted_eclipse_depths} list the transit and eclipse depths we used in the retrieval, along with the papers they were taken from.

\begin{table}[ht]
  \centering
  \caption{Adopted transit depths. Sources: (1) \citealt{pont_2013}; (2) \citealt{mccullough_2014}}
  \begin{tabular}{C C C C C}
  \hline
  	\textrm{$\lambda_{min}$(\um)} & \textrm{$\lambda_{max}$(\um)} & \textrm{Depth (ppm)} & \textrm{Error (ppm)} & \textrm{Source}\\
    \hline
0.32 & 0.37 & 24999 & 101 & 1\\
0.37 & 0.42 & 24809 & 50 & 1\\
0.42 & 0.47 & 24706 & 44 & 1\\
0.47 & 0.52 & 24633 & 41 & 1\\
0.52 & 0.56 & 24542 & 41 & 1\\
0.56 & 0.58 & 24473 & 44 & 1\\
0.58 & 0.592 & 24455 & 84 & 1\\
0.592 & 0.604 & 24433 & 69 & 1\\
0.604 & 0.615 & 24389 & 112 & 1\\
0.615 & 0.626 & 24336 & 84 & 1\\
0.626 & 0.638 & 24367 & 37 & 1\\
0.65 & 0.7 & 24289 & 34 & 1\\
0.7 & 0.75 & 24249 & 34 & 1\\
0.75 & 0.8 & 24292 & 37 & 1\\
0.8 & 0.85 & 24186 & 37 & 1\\
0.85 & 0.9 & 24190 & 40 & 1\\
0.9 & 0.95 & 24168 & 40 & 1\\
0.95 & 1.0 & 24186 & 50 & 1\\
1.0 & 1.17 & 24062 & 68 & 1\\
1.1184 & 1.1374 & 23962 & 73 & 2\\
1.1372 & 1.1562 & 24047 & 67 & 2\\
1.156 & 1.175 & 24078 & 105 & 2\\
1.1748 & 1.1938 & 24035 & 87 & 2\\
1.1936 & 1.2126 & 23961 & 80 & 2\\
1.2123 & 1.2313 & 23955 & 70 & 2\\
1.2311 & 1.2501 & 23884 & 56 & 2\\
1.2499 & 1.2689 & 24000 & 62 & 2\\
1.2687 & 1.2877 & 23863 & 61 & 2\\
1.2875 & 1.3065 & 23987 & 69 & 2\\
1.3062 & 1.3252 & 23961 & 60 & 2\\
1.325 & 1.344 & 23982 & 66 & 2\\
1.3438 & 1.3628 & 24134 & 55 & 2\\
1.3626 & 1.3816 & 24149 & 61 & 2\\
1.3814 & 1.4004 & 24091 & 63 & 2\\
1.4001 & 1.4191 & 24215 & 77 & 2\\
1.4189 & 1.4379 & 24199 & 64 & 2\\
1.4377 & 1.4567 & 24108 & 71 & 2\\
1.4565 & 1.4755 & 24018 & 67 & 2\\
1.4752 & 1.4942 & 24188 & 75 & 2\\
1.494 & 1.513 & 23941 & 62 & 2\\
1.5128 & 1.5318 & 24097 & 61 & 2\\
1.5316 & 1.5506 & 24002 & 62 & 2\\
1.5504 & 1.5694 & 24010 & 72 & 2\\
1.5691 & 1.5881 & 24100 & 87 & 2\\
1.5879 & 1.6069 & 23963 & 75 & 2\\
1.6067 & 1.6257 & 23916 & 98 & 2\\
1.6255 & 1.6445 & 24062 & 84 & 2\\
3.2 & 3.9 & 24047 & 84 & 1\\
4.0 & 5.0 & 24155 & 109 & 1\\
5.0 & 6.4 & 23951 & 207 & 1\\
6.4 & 9.3 & 24056 & 105 & 1\\
23.5 & 24.5 & 23898 & 291 & 1\\
    \hline
  \end{tabular}
  \label{table:adopted_transit_depths}
\end{table}

\begin{table*}[ht]
  \centering
  \caption{Adopted eclipse depths. Sources: (1) \citealt{crouzet_2014}; (2) \citealt{kilpatrick_2020}; (3) \citealt{charbonneau_2008}; (4) \citealt{agol_2010}}
  \begin{tabular}{C C C C C}
  \hline
  	\textrm{$\lambda_{min}$(\um)} & \textrm{$\lambda_{max}$(\um)} & \textrm{Depth (ppm)} & \textrm{Error (ppm)} & \textrm{Source}\\
    \hline
1.1184 & 1.1374 & 0 & 47 & 1\\
1.1372 & 1.1562 & 78 & 50 & 1\\
1.156 & 1.175 & 124 & 45 & 1\\
1.1748 & 1.1938 & 93 & 44 & 1\\
1.1936 & 1.2126 & 89 & 43 & 1\\
1.2123 & 1.2313 & 51 & 50 & 1\\
1.2311 & 1.2501 & 64 & 42 & 1\\
1.2499 & 1.2689 & 99 & 42 & 1\\
1.2687 & 1.2877 & 80 & 42 & 1\\
1.2874 & 1.3064 & 149 & 41 & 1\\
1.3062 & 1.3252 & 127 & 41 & 1\\
1.325 & 1.344 & 108 & 41 & 1\\
1.3438 & 1.3628 & 75 & 41 & 1\\
1.3626 & 1.3816 & 69 & 41 & 1\\
1.3813 & 1.4003 & 56 & 42 & 1\\
1.4001 & 1.4191 & 104 & 42 & 1\\
1.4189 & 1.4379 & 94 & 42 & 1\\
1.4377 & 1.4567 & 67 & 42 & 1\\
1.4565 & 1.4755 & 32 & 43 & 1\\
1.4753 & 1.4943 & 105 & 43 & 1\\
1.494 & 1.513 & 228 & 43 & 1\\
1.5128 & 1.5318 & 211 & 43 & 1\\
1.5316 & 1.5506 & 143 & 44 & 1\\
1.5504 & 1.5694 & 99 & 45 & 1\\
1.5691 & 1.5881 & 135 & 45 & 1\\
1.5879 & 1.6069 & 80 & 46 & 1\\
1.6067 & 1.6257 & 32 & 46 & 1\\
1.6255 & 1.6445 & 110 & 74 & 1\\
3.2 & 4.0 & 1481 & 34 & 2\\
4.0 & 5.0 & 1827 & 22 & 2\\
5.1 & 6.3 & 3100 & 340 & 3\\
6.6 & 9.0 & 3440 & 36 & 4\\
13.5 & 18.5 & 5190 & 220 & 3\\
20.8 & 26.1 & 5980 & 380 & 3\\
    \hline
  \end{tabular}
  \label{table:adopted_eclipse_depths}
\end{table*}

\begin{table*}[ht]
  \centering
  \caption{ExoMol line lists used in \platon v5}
  \begin{tabular}{|p{1.3cm}|p{2cm}|p{2cm}|p{1.2cm}|p{4cm}|}
  \hline
  	Molecule & List name & N\textsubscript{lines} & T\textsubscript{max}(K) & Reference\\
    \hline
    C$_2$H$_4$ & MaYTY & 49,673,223,799* & 700 & \cite{mant_2018}\\
    CO & Li2015 & 125,496* & 9000 & \cite{li_2015}\\
    H$_2$CO & AYTY & 12,648,694,479* & 1500* & \cite{al-refaie_2015}\\
    H$_2$S & AYT2 & 115,623,180* & 2000* & \cite{azzam_2016}\\
    H$_2$O & POKAZATEL & 5,550,587,708* & $\infty^b$* & \cite{polyansky_2018}\\
    HCl & Yueqi & 2588 & ? & \cite{li_2013}\\
    HCN & Harris & 34,418,408* & ? & \cite{barber_2013}\\
    MgH & MoLLIST & 14,179* & ? & \cite{gharibnezhad_2013}\\
    NH$_3$ & CoYuTe & 1,135,240,003* & 1500* & \cite{coles_2019}\\
    NO & NOname & 2,280,366* & ? & \cite{wong_2017}\\
    OH & MoLLIST & 54,276 & ? & \cite{brooke_2016}\\
    PH$_3$ & SAlTY & 16,931,647,841* & 1500* & \cite{sousa-silva_2014}\\
    SH & SNaSH & 81,348* & 5000 & \cite{yurchenko_2018b}\\
    SiH & SiGHTLY & 1,724,841* & 5000 & \cite{yurchenko_2017}\\
    SiO & EBJT & 254,675* & 9000 & \cite{barton_2013}\\
    SO$_2$ & ExoAmes & 1,402,257,689* & 2000 & \cite{underwood_2016}\\
    TiO & ToTo & 295,086,011$^a$ & 5000 & \cite{mckemmish_2019}\\
    VO & VOMYT & 277,131,624 & 5000 & \cite{mckemmish_2016}\\
      \hline
  \end{tabular}
      \tablecomments{$^{(a)}$All 5 isotopologues combined.  Each isotopologue has 58-60 million lines.}
      \tablecomments{$^{(b)}$This line list is complete}
      \tablecomments{$^*$These numbers disagree with those in the ExoMol database's def files}
  \label{table:exomol_line_lists}
\end{table*}

\bibliographystyle{apj} \bibliography{main}
\end{document}